\documentclass[aps,
		prd,
		reprint,
		onecolumn,
		superscriptaddress,
		shortbibliography,
		nofootinbib,
		floatfix,
		notitlepage
		]{revtex4-1}

\usepackage{amsmath,amssymb,amsfonts}
\usepackage[dvipsnames]{xcolor}
\usepackage{hyperref}
\hypersetup{colorlinks=true,citecolor=Cyan,urlcolor=Red}
\usepackage{graphicx}

\usepackage{slashed}

\newcommand{\be}{\begin{equation}}
\newcommand{\ee}{\end{equation}}
\newcommand{\bi}{\begin{itemize}}
\newcommand{\ei}{\end{itemize}}
\newcommand{\bea}{\begin{eqnarray}}
\newcommand{\eea}{\end{eqnarray}}
\newcommand{\ket}[1]{|\,#1\,\rangle}          
\newcommand{\ud}{\mathrm{d}}		

\newcommand{\LCm}{{\scriptscriptstyle -}} 
\newcommand{\LCp}{{\scriptscriptstyle +}}

\newcommand{\LCperp}{{\scriptscriptstyle \perp}}

\newcommand\redsout{\bgroup\markoverwith{\textcolor{red}{\rule[0.5ex]{2pt}{0.4pt}}}\ULon}
\usepackage[normalem]{ulem}
\makeatletter
\newcommand\wave{\bgroup \markoverwith{\textcolor{red}{\lower3.5\p@\hbox{\sixly \char58}}}\ULon}
\makeatother

\begin{document}

\title{Scattering in strong electromagnetic  fields: transverse size effects in tBLFQ}
\author{Bolun Hu}
\email{hubolun@impcas.ac.cn}
\affiliation{School of Physical Science and Technology, Lanzhou University, Lanzhou 730000, China}
\affiliation{Institute of Modern Physics, Chinese Academy of Sciences, Lanzhou 730000, China}
\affiliation{School of Nuclear Science and Technology, University of Chinese Academy of Sciences, Beijing 100049, China}
\author{Anton Ilderton}
\email{anton.ilderton@plymouth.ac.uk}
\affiliation{Centre for Mathematical Sciences, University of Plymouth, PL48AA, United Kingdom}

\author{Xingbo Zhao}
\email{xbzhao@impcas.ac.cn}
\affiliation{Institute of Modern Physics, Chinese Academy of Sciences, Lanzhou 730000, China}
\affiliation{School of Nuclear Science and Technology, University of Chinese Academy of Sciences, Beijing 100049, China}

\begin{abstract}	
   The framework of `time-dependent basis light-front quantisation' (tBLFQ) offers a non-perturbative approach to scattering problems in external fields, based on Fock space truncation. Here we extend tBLFQ to include spatio-temporal field inhomogeneities in multiple spacetime directions. This extension is necessary for the proper modelling of e.g.~intense laser fields. We focus on the example of nonlinear Compton scattering of an electron on an axicon-type laser, with an emphasis on the transverse structure of the beam. We analyse the impact of field intensity and particle energy, as well as basis truncation effects, on the radiation spectrum of the scattered electron.
\end{abstract}

\maketitle

\section{Introduction}

The theoretical predictions of perturbative QED, such as electron $g-2$~\cite{Hanneke:2008tm,Karshenboim:2005iy} and the Lamb shift~\cite{Miller:2012opa}, have been experimentally verified to extreme precision. However, there are many problems for which standard perturbative methods are insufficient, such as bound states~\cite{Hoyer:2018hdj}, or break down completely, such as strongly coupled systems. Thus other methods are required to access \textit{non}-perturbative physics.

In the context of external field problems, perhaps the most famous non-perturbative effect is the spontaneous conversion of a sufficiently strong electric field into electron-positron pairs, or Schwinger pair production, the probability of which is a non-perturbative function of the coupling~\cite{Schwinger:1951nm}. It may be possible to observe Schwinger pair creation using the ultra-intense electromagnetic fields of future lasers~\cite{Bulanov:2010ei,Gonoskov:2013ada}, and there is currently an international effort to develop both the theoretical and experimental tools necessary to investigate this and other quantum effects in strong (external) fields, such as vacuum birefringence; for a review see~\cite{King:2015tba}.

When investigating non-perturbative phenomena, exactly solvable systems offer intuition, but one often turns to numerical schemes which do not rely on perturbative approximations. The lattice is a well-known example, and real-time lattice techniques can now be used to analyse e.g.~string breaking and the Schwinger effect~\cite{Spitz:2018eps}. Within laser-plasma physics, a common tool is the simulation of processes using Particle-In-Cell (PIC) codes, in which Monte-Carlo routines based on QED calculations are included to simulate (tree level) quantum effects; for a review see~\cite{Gonoskov:2014mda}. The advantages of such approaches are that extremely complicated electron-positron-photon interactions can be simulated, with arbitrarily complex laser fields. The disadvantage, though, is that such methods are firmly rooted in classical physics, and the approximations behind the inclusion of quantum effects are known to break down at both low~\cite{Harvey:2014qla,DiPiazza:2017raw,Ilderton:2018nws}, and high energy~\cite{Podszus:2018hnz,Ilderton:2019kqp}. Here we consider a different approach; it is not our goal to compete with e.g.~PIC schemes, but to complement them with an alternative, and fully quantum mechanical, framework.

Any numerical implementation of QFT requires a cutoff in order to render problems finite-dimensional. Because the set of all products of one-particle states gives a possible basis of Hilbert space~\cite{Weinberg:1995mt}, one can try to solve problems using a finite truncation of this basis, corresponding to a truncation in particle number~\cite{Tamm:1945qv,Dancoff:1950ud}. This is the main idea of basis light-front quantisation (BLFQ)~\cite{Vary:2009gt}; one solves the Schr\"odinger equation in a truncated Fock space, rather than a perturbative expansion. Thus BLFQ is a non-perturbative approach. An extension of this framework which is suitable for investigating e.g.~QFT scattering processes in \textit{time-dependent} background fields is time-dependent basis light-front quantisation (tBLFQ)~\cite{Zhao:2013cma}.

Previously the emission of photons from an electron scattering off an intense background field (called nonlinear Compton scattering)~\cite{Nikishov:1964zza} has been studied in tBLFQ, but in the approximation that the background has essentially univariate spacetime dependence~\cite{Zhao:2013cma,Zhao:2013jia}. However, intense laser fields for example are strongly spatially focussed, and so further progress in high-intensity laser-matter interactions requires the consideration of complex background field geometries.  Motivated by this, and by a desire to improve the overall scope of tBLFQ, we here extend the framework to include effects due to multi-dimensional field inhomogeneities.

This paper is organised as follows. We review the main ideas of BLFQ and tBLFQ in Sec.~\ref{SECT:BACKGROUND}. In Sec.~\ref{SECT:SINGLEELECTRON} we use tBLFQ to study the simpler system of a single electron in a static, but position-dependent, background field (what will be the `centre-of-mass'  motion of the system). In Sec.~\ref{SECT:PHOTONS} we reintroduce the dynamical photon fields and study photon emission from the electron; observables are defined and their time-evolution is studied. In Sec.~\ref{SECT:TIME}, we add a time dependence to the background. We present our conclusions and outlook in Sec.~\ref{SECT:OUTLOOK}. The appendices include conventions and other details behind the calculations in the text.

\section{Background}\label{SECT:BACKGROUND}
%
In the light-front formalism, field theories are quantised at fixed light-front time $x^{\LCp}=x^{0}+x^{3}$, following~\cite{Dirac:1949cp}, rather than at fixed $x^{0}$ as in the equal-time formalism; see~\cite{Brodsky:1997de,Heinzl:2000ht} for reviews. The remaining coordinates are $x^\LCm = x^0 - x^3$ (longitudinal) and $x^\LCperp = \{x^1,x^2\}$ (transverse).

Due to the larger numbers of \textit{kinematic} Lorentz generators, over \textit{dynamical} generators, in light-front coordinates, wavefunctions in one frame can be more easily mapped to those in other frames. Furthermore, light-front kinematics forbids massive particles from contributing to the vacuum, therefore making the trivial Fock vacuum the vacuum of the full theory. (This holds only if one neglects zero-mode contributions, which it is not always safe to do~\cite{Mccartor:1988bc,Hornbostel:1988ne,Bogolyubov:1990kw,Ji:1995ft,McCartor:1996nj,Tomaras:2001vs,Heinzl:2003jy,Mannheim:2019lss}.)

One consequence of this is that the physical cases of interest here are easily expandable in simple Fock modes. For example, if the vacuum received its most significant contribution from some higher Fock sector, take for example $|\,e^{+}e^{-}\,\rangle$, then the Fock expansion of the physical electron would `begin' with $\ket{e}_{\rm phys} \sim \ket{e e e^{+}}$. The triviality of the vacuum means though that the first contributions take the form $|\,e\,\rangle_{\rm phys}=a|\,e\,\rangle+b|\,e\gamma\,\rangle$. It is the simplicity of the Fock expansion that makes the Hamiltonian formalism feasible in QFT.


\subsection{Time-evolution in light-front field theory}\label{SECT:SEQUATION}
In the Hamiltonian formalism of QFT, the time-evolution of a system is governed by the Schr\"odinger equation, which in light-front quantisation takes the form
\begin{equation}
   i\frac{\partial}{\partial x^{\LCp}}|\,\Psi;x^{\LCp}\,\rangle_{\mathrm S}=\frac{1}{2}P_{\mathrm S}^{\LCm}(x^{\LCp})|\,\Psi;x^{\LCp}\,\rangle_{\mathrm S}\;,
     \label{}
\end{equation}
in which the Schr\"odinger picture Hamiltonian $P^{\LCm}_{\mathrm S}$ contains two parts, 
\begin{equation}
   P_{\mathrm S}^{\LCm}(x^{\LCp})=P^{\LCm}_{\mathrm{QED}}+V_{\mathrm S}(x^{\LCp}) \;,
  \label{eqn:hamiltonian}
\end{equation}
where $P^{\LCm}_{\mathrm{QED}}$ is the full light-front Hamiltonian of, here, QED, and $V_{\mathrm S}$ consists of additional interaction terms introduced by a background field, to be specified explicitly below. If the effects introduced by this background are the primary source of interest, then it is convenient to work in an interaction picture in which the `free' states are eigenstates of the full QED Hamiltonian $P^{\LCm}_{\mathrm{QED}}$, and where the only nontrivial time-evolution is induced by the new interaction $V_{\mathrm S}$. It is of course not possible to do this exactly for QED, and hence a numerical approximation will be introduced below. In principle, though, in this interaction picture, the Schr\"odinger equation becomes
\begin{equation}
  i\frac{\partial}{\partial x^{\LCp}}|\,\Psi;x^{\LCp}\,\rangle_{\rm I}=\frac{1}{2}V_{\rm I}(x^{\LCp})|\,\Psi;x^{\LCp}\,\rangle_{\rm I}\;, 
  \label{eqn:interaction}
\end{equation}
the formal solution to which is
\begin{equation}
   |\,\Psi;x^{\LCp}\,\rangle_{\rm I}=\mathcal{T}_{\LCp}\exp{\left(-\frac{i}{2}\int^{x^{\LCp}}_{0}dx'^{\LCp}V_{\rm I}(x'^{\LCp})\right)}|\,\Psi;0\,\rangle_{\rm I}\;.
  \label{eqn:time}
\end{equation}
States and operators in the two pictures are as usual related by
\begin{equation}
   |\,\Psi;x^{\LCp}\,\rangle_{\rm I}=e^{\frac{i}{2}P^{\LCm}_{\mathrm{QED}}x^{\LCp}}|\,\Psi;x^{\LCp}\,\rangle_{\rm S}\;, \qquad   A_{\rm I}(x^{\LCp})=e^{\frac{i}{2}P^{\LCm}_{\mathrm{QED}}x^{\LCp}}A_{\rm S}(x^{\LCp})e^{-\frac{i}{2}P^{\LCm}_{\mathrm{QED}}x^{\LCp}} \;.
   \label{eqn:interactionpictureoperator}
\end{equation}
In this interaction picture, we work in a basis of eigenstates of $P_{\rm QED}^{\LCm}$. This basis, called the `tBLFQ basis' $|\,\beta\,\rangle$ (more details will be presented in the next section), will simplify the operator exponentials in the definition of the operators in Eq.~(\ref{eqn:interactionpictureoperator}). In this basis the matrix elements of the interaction Hamiltonian $V_{\rm I}$ become 
\begin{equation}
\langle\,\beta'\,|\,V_{\rm I}\,|\,\beta\,\rangle=\langle\,\beta'\,|\,V_{\rm S}\,|\,\beta\,\rangle\mathrm{exp}\left[\frac{i}{2}(P^{\LCm}_{\beta'}-P^{\LCm}_{\beta})x^{\LCp})\right].
  \label{eqn:phasefactor}
\end{equation}
It is then straightforward to evolve the quantum state according to Eq.~(\ref{eqn:time}), by decomposing the time-evolution operator into many small steps of light-front time $x^{\LCp}$ with a step size $\delta x^{\LCp}$
\begin{equation}
   \begin{split}
      \mathcal{T}_{\LCp}\exp{\left(-\frac{i}{2}\int^{x^{\LCp}}_{0}dx'^{\LCp}V_{\rm I}(x'^{\LCp})\right)}\to \left[1-\frac{i}{2}V_{\rm I}(x^{\LCp}_{n})\delta x^{\LCp}\right]\cdots\left[1-\frac{i}{2}V_{\rm I}(x^{\LCp}_{1})\delta x^{\LCp}\right]\;.
      \label{eqn:discreteizedtime}
   \end{split}
\end{equation}
This discretisation is the Euler scheme; it is however not usually a good choice. The most significant shortcoming of this scheme is its poor stability (which means that the norm of the state vector $|\,\Psi;x^{\LCp}\,\rangle$ changes as time evolves). Instead, we will adopt the second-order difference scheme MSD2, which has been proved to have better stability~\cite{askar1978askar}.  This scheme relates the state at $x^{\LCp}+\delta x^{\LCp}$ to those at both $x^{\LCp}$ and $x^{\LCp}-\delta x^{\LCp}$ (rather than just that at $x^{\LCp}$ as in the Euler scheme), thus:
\begin{equation}
   |\,\Psi;x^{\LCp}+\delta x^{\LCp}\,\rangle_{\rm I}=|\,\Psi;x^{\LCp}-\delta x^{\LCp}\,\rangle_{\rm I}+(e^{-iV_{\rm I}\delta x^{\LCp}/2} -e^{iV_{\rm I}\delta x^{\LCp}/2})|\,\Psi;x^{\LCp}\,\rangle_{\rm I}\\
   \approx |\,\Psi;x^{\LCp}-\delta x^{\LCp}\,\rangle_{\rm I}-iV_{\rm I}|\,\Psi;x^{\LCp}\,\rangle_{\rm I}\;.
   \label{eq:msd2}
\end{equation}
Once the quantum state at each time step is obtained, it is straightforward to construct observables from it.
\subsection{Basis light-front quantisation}
As mentioned above, it is convenient to evolve the system in the tBLFQ basis, which by definition comprises the eigenstates of $P^{\LCm}_{\rm QED}$. In such a basis, the complicated exponentials of the operator in Eq.~(\ref{eqn:interactionpictureoperator}) reduce to phase factors, which greatly simplifies the computation. The eigenstates themselves are constructed in BLFQ, which is a Hamiltonian formalism incorporating the advantages of light-front dynamics. The main idea is to solve, numerically and in a Fock space truncation, the time-independent Schr\"odinger equation
\begin{equation}
   P^{\LCm}_{\mathrm{QED}}|\,\beta\,\rangle=P^{\LCm}_{\beta}|\,\beta\,\rangle \;.
   \label{eigenequation}
\end{equation}
BLFQ has the advantage of being able to solve bound state problems involving positronium~\cite{Wiecki:2015xxa} and hadron structures~\cite{Li:2015iaw,Li:2015zda,Karmanov:2016yzu,Chen:2016dlk,Li:2017mlw,Adhikari:2018umb,Tang:2018myz,Jia:2018ary,Jia:2018hxd,Lan:2019vui,Lan:2019rba,Du:2019qsz,Mondal:2019yph,Lan:2019img,Xu:2019xhk,Du:2019ips}. In this paper we follow previous work on the \textit{physical} electron eigenstates in BLFQ~\cite{Zhao:2014xaa}. A sector-dependent renormalisation~\cite{Karmanov:2008br,Karmanov:2012aj} of the electron mass is performed, improving over~\cite{Zhao:2013cma}. We review the main steps here; more details may be found in~\cite{{Zhao:2013cma}}.

 Eq.~(\ref{eigenequation}) is infinite-dimensional. To reduce the equation to a finite-dimensional problem a truncation of the basis should be implemented. We are in this paper interested in transitions between the physical electron state and its excitations, which are scattering states, due to the interaction introduced by a background field. For simplicity we retain only the first two Fock sectors $|\,e\,\rangle$ and $|\,e\gamma\,\rangle$, which are enough to give a description of the photon emitted from the electron excited by the background field. In this approximation, the physical electron and photon states have the form, with $\#$ indicating some coefficients,
 \be
 	\ket{e}_\text{phys} = \# \ket{e} + \# \ket{e\gamma} \;, \qquad \ket{\gamma}_\text{phys} = \# \ket{\gamma} \;.
\ee
This truncation of the Fock space implicitly assumes that higher Fock sectors give decreasing contributions to the low-lying eigenstates, in which we are most interested; one motivation for this is the success of perturbation theory in QED, and further details of this approximation to the physical states will be discussed in Sec.~\ref{SECT:OBSERVABLE}.

We now characterise the single-particle Fock sector states themselves. These carry four quantum numbers, ultimately corresponding to three momentum components and a spin or helicity.  The first quantum number $k$ labels the longitudinal momentum $p^{\LCp}$ of the particle. We compactify the longitudinal direction $x^{\LCm}$ on a circle of length $2L$ and impose (anti)periodic boundary conditions on (fermions) bosons. The longitudinal momentum $p^{\LCp}$ therefore takes the discrete values
 \begin{equation}
    p^{\LCp}=\frac{2 \pi}{L} k\;,
    \label{eqn:disc_longitudinal}
 \end{equation}
 where the dimensionless quantity \(k=1,2,3, \ldots\) for bosons (neglecting the zero mode) and \(k=\frac{1}{2}, \frac{3}{2}, \frac{5}{2}, \ldots\) for fermions.  The next two quantum numbers $n$ and $m$ are those of a 2D harmonic oscillator (2D-HO) in the transverse plane, of mass $M$ and frequency $\Omega$. These numbers therefore encode the transverse momenta. The 2D-HO eigenstates have the corresponding eigenvalues
 \begin{equation}
    E_{n, m}=(2 n+|\,m|\,+1) \Omega\;.
    \label{hoenergy}
 \end{equation}
Note though that the only parameter of the 2D-HO that enters the eigenstates is the scale parameter $b:=\sqrt{M\Omega}$; for details see Appendix~\ref{SECT:HOBASIS}.  The fourth and final quantum number $\lambda$ is the light-front helicity of the particle. We write  $\bar{\alpha}=\{k, n, m, \lambda\}$ as a shorthand for the four quantum numbers and define the corresponding single-particle state as $|\,\bar \alpha\,\rangle$~\cite{Zhao:2013cma}.  These states are chosen so as to preserve as many symmetries of the Hamiltonian as possible, and thus simplify calculations in BLFQ.  Compared with a standard plane-wave basis, the 2D-HO basis preserves rotational symmetry in the transverse plane, even in the finitely truncated BLFQ basis. The shortcoming of the 2D-HO basis states is that they are not eigenstates of momentum and consequently it becomes difficult to separate the relative motion from the centre-of-mass motion. We will however identify a method to resolve this, in Sec.~\ref{SECT:RESULT}. For more details of the symmetries of the QED Hamiltonian and the basis, see \cite{Zhao:2013cma,Li:2013cga}. To construct $N$-particle states $|\, \bar\alpha_{N}\,\rangle$, we simply take the direct product of single-particle states $|\,\bar\alpha_{N} \,\rangle=\otimes |\, \bar{\alpha} \,\rangle$. The working BLFQ basis $|\,\alpha\,\rangle$, is then the direct sum of all single- and multi-particle states retained within after the Fock truncation.

Even with the restriction in Fock number, the parameters in $\bar\alpha$ (aside from the helicity) are still unbounded and must also be truncated. To impose this, we introduce two parameters $K_{\rm total}$ and $N_{\rm max}$ with which to truncate the BLFQ basis in the longitudinal and transverse directions respectively.  For the longitudinal degrees of freedom, $K_{\rm total}$ is defined by
 \begin{equation}
    K_{\rm total}=\sum_i k_{i}\;,
    \label{}
 \end{equation}
 where the sum runs over all single-particle states. If the total longitudinal momentum is conserved, as is the case in  this paper, then only a single $K_{\rm total}$ is enough to describe the system. Furthermore, because the light-front wavefunctions are functions of longitudinal momentum fractions $\mathrm x_{i}=k_{i}/K_{\rm total}$, rather than longitudinal momenta themselves, larger $K_{\rm total}$ only provides a finer description of the system.
 
In the transverse plane, we define the total transverse quantum number for a BLFQ basis state \(|\,\alpha\,\rangle\) as
 $$
 N_{\alpha}=\sum_{i} 2 n_{i}+\left|\,m_{i}\right|\,+1\;,
 $$
 where the sum again runs over all particles in the state. All the retained basis states satisfy
 \begin{equation}
    N_{\alpha} \leq N_{\max }\;.
    \label{}
 \end{equation}
 Physically, $N_{\rm max}$ limits the total `energy' of 2D-HO states summed over all particles. $N_{\rm max}$ is specified globally across all Fock sectors to ensure that the transverse motion in different Fock sectors is truncated at the same energy. As a result, $N_{\rm max}$ determines both the UV and IR cutoffs for the transverse basis; for details, see~\cite{Coon:2012ab,Furnstahl:2012qg}.

 At this stage we have a finite basis in which to work. Working in the BLFQ basis $\ket{\alpha}$ we can diagonalise the Hamiltonian in order to find the QED eigenstates $|\,\beta\,\rangle$ and eigenvalues $P^{\LCm}_{\beta}$ satisfying the time-independent Schr\"odinger equation~(\ref{eigenequation}). (It is straightforward to obtain the matrix elements in the BLFQ basis by using the commutation relations of the creation and annihilation operators of the Fock states; see the Appendix, Eq.~(\ref{eqn:dis_commutation})). From there, matrix elements $\langle\,\beta'\,|\,V_{\rm I}\,|\,\beta\,\rangle$ as in Eq.~(\ref{eqn:phasefactor}) are easily calculated in terms of the wavefunction $\langle\, \alpha\,|\,\beta\,\rangle$; for more details see Appendix \ref{SECT:INTMATRIX}.
\section{Particle in an external field}\label{SECT:SINGLEELECTRON}

In this section, we consider the simplest limit of our system, namely single particles in an external field. In the language of tBLFQ, this means that we retain only the single-electron sector $|\,e\,\rangle$ in our calculations, neglecting the electron-photon sector $|\,e\gamma\,\rangle$ (and others). In doing so, we effectively turn off dynamical photon generation and absorption. These will be reinstated in Sec.~\ref{SECT:PHOTONS}. 

\begin{figure}[t!]
   \centering
   \includegraphics[height=6cm]{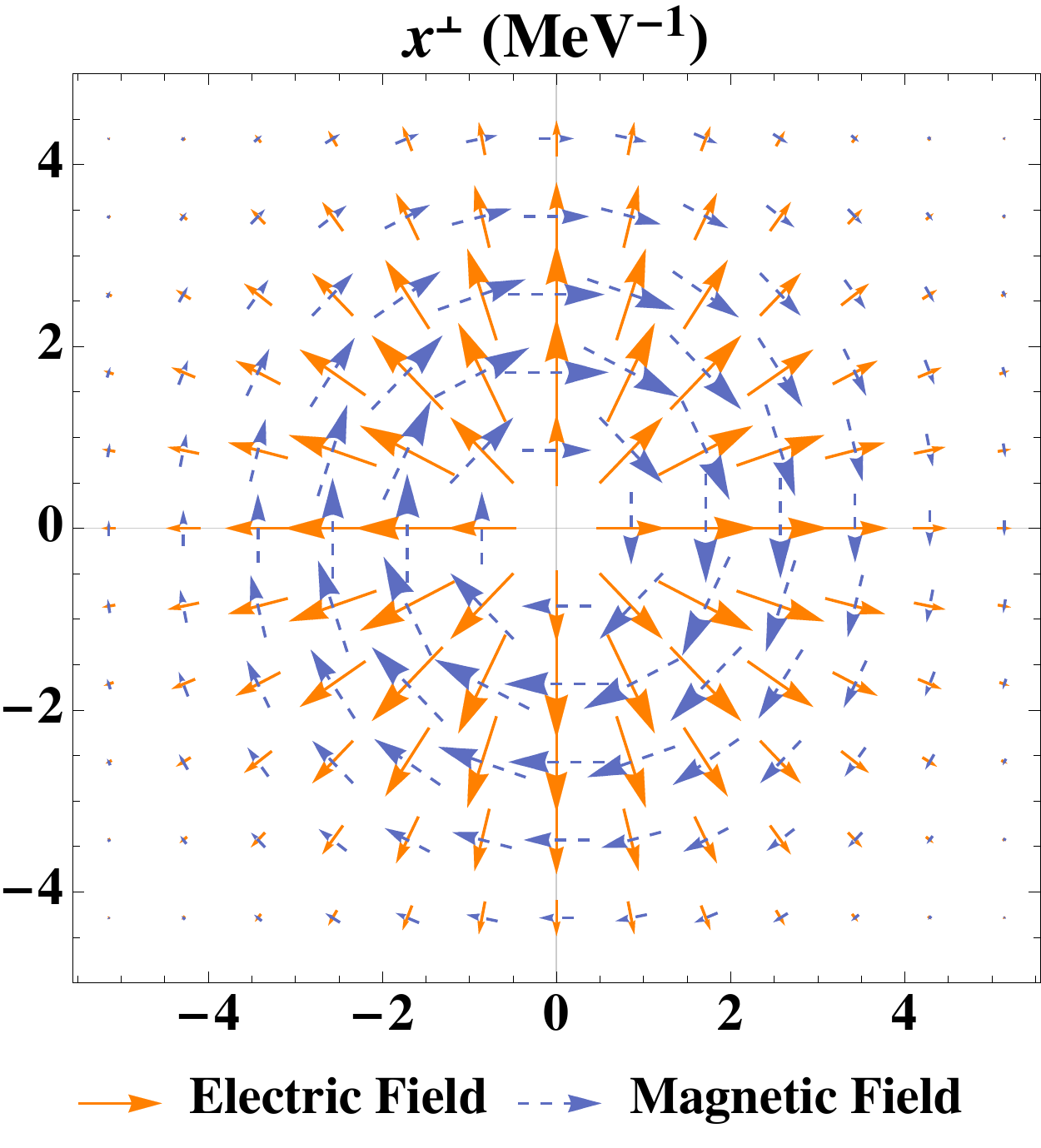}
   \includegraphics[height=5.5cm]{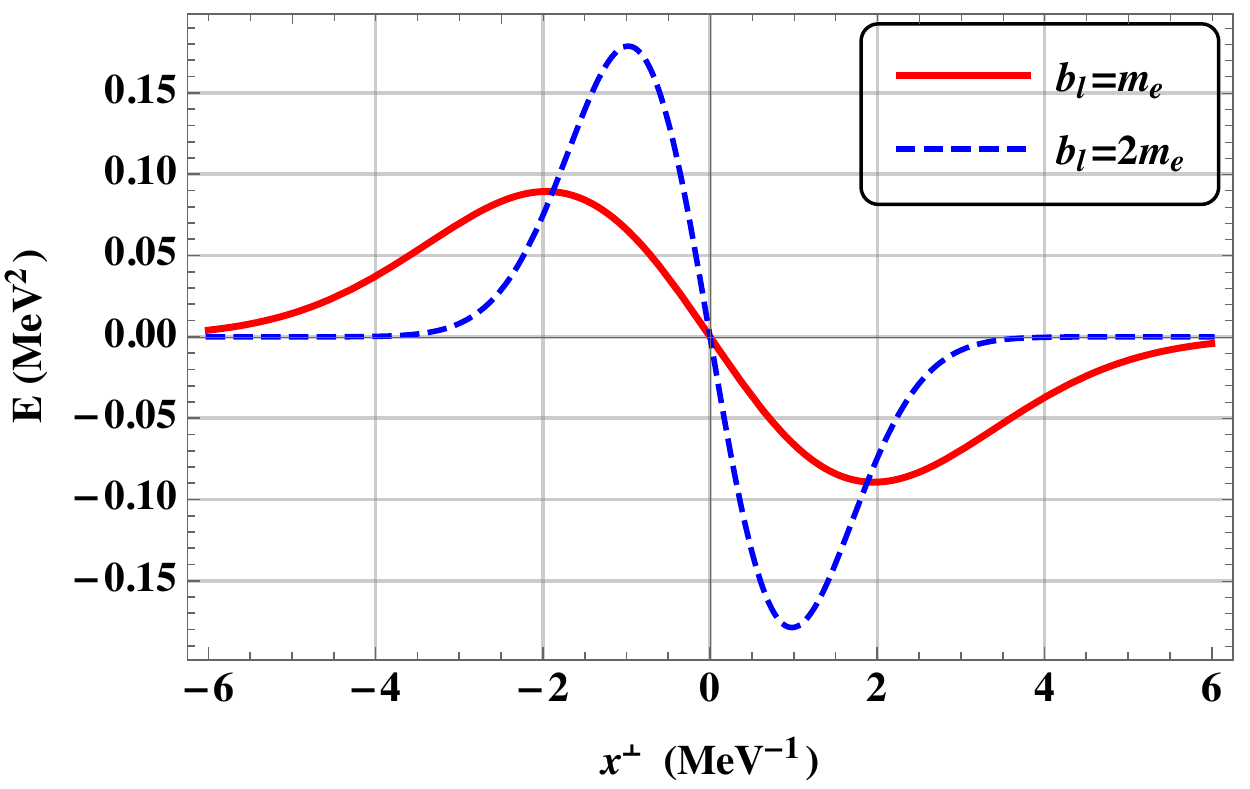}
   \caption{\label{eandb} \textbf{Left}: Transverse electric and magnetic fields of the background, shown as solid and dashed lines respectively, as a function of the transverse coordinates. Parameters: $a_{0}=1, b_{l}=m_{e}$. The two fields are of the same magnitude and are perpendicular at every point.   \textbf{Right}: The transverse electric field for different widths $b_l$. As $b_l$ increases the field becomes narrower in position space, but reaches higher peak values.}
\end{figure}

A general background field generates four new interaction terms in the full light-front Hamiltonian of QED, given in full in Eq.~(\ref{eqn:fullhamiltonian}). Three of these vanish if we can represent the background by a potential having only a single, longitudinal, component $\mathcal{A}^\LCm$. We make this simplying assumption from here on; note that there remains a large amount of freedom in the choice of the background field. Our choice is motivated by the desire to go beyond previous tBLFQ results \cite{Zhao:2013cma} and include field dependence on the transverse coordinates $x^\LCperp$. Consider then an `axicon' laser beam propagating in the $z$-direction. The beam has a radially polarised electric field and an azimuthally polarised magnetic field, both transverse to the propagation direction. This implies the vanishing of the electric field on the symmetry, or $z$, axis~\cite{McDonald:2000b,Heinzl:2017zsr}.  Typically one takes the transverse fields to go like $\sim |x^\LCperp| \exp -|x^\LCperp|^2$. Axicon beams also have longitudinal fields (i.e.~fields pointing in the $z$-direction) with essentially the same temporal (or $x^\LCp$) dependence as the transverse fields, but which are suppressed relative to the transverse by a small focussing parameter. Thus a toy model of an axicon beam can be given by ignoring the longitudinal fields, and taking the potential to be
\begin{equation}
   e\mathcal{A}^{\mu}(x)= \delta^{\mu}_{\LCm} \frac{m_e a_0}{\sqrt{\pi}} \exp \bigg(-\frac{1}{2}b_l^2 x^\LCperp x^\LCperp\bigg) f(x^\LCp)\;,
   \label{BG-AI}
\end{equation}
in which $f(x^\LCp)$ encodes the time dependence of the field; $a_0$ is a dimensionless amplitude, and $b_l$ is a width. We will set $f\equiv 1$ to begin with, re-introducing explicit time dependence in Sec.~\ref{SECT:TIME}. Fig.~\ref{eandb} shows the resulting, transverse, electric and magnetic fields. These are radially and azimuthally polarised respectively (as desired), orthogonal and of equal magnitude at every point. The transverse field profile as a function of $b_l$ is also shown.

A further motivation for this choice of background is that its symmetries match well with those of the BLFQ basis; the potential (\ref{BG-AI}) is proportional to the lowest order 2D-HO eigenstate $\Phi_{00}$ in coordinate space (recall $f\equiv 1$ for now and see Appendix \ref{SECT:HOBASIS} for details on the basis), 
\begin{equation}
   e\mathcal{A}^{\mu}(x)= \delta^{\mu}_{\LCm} \frac{m_ea_0}{b_l} \, \Phi_{00}^{b_{l}}(x^{\LCperp}) \;.
   \label{eqn:backgroundfield}
\end{equation}
The classical physics of a particle in the model axicon field above is straightforward. Consider a particle near the symmetry axis. Depending on its charge, or equivalently the sign of $a_0$, the particle is either repelled from or attracted to the axis. In the former case, see the right-hand panel of Fig.~\ref{eandb}, the particle will be accelerated into the weak-field region and then drift outward at some acquired velocity. Particles of opposite charge will be attracted to the symmetry axis, which is also a weak field region, but will overshoot, and may then be attracted back again, oscillating around the axis. This intuition will help us analyse the results of the tBLFQ calculation, below.

\subsection{Time-evolution in tBLFQ}
We now give the first tBLFQ calculation. We begin with a single-electron state and consider its evolution in the axicon background above. We take the longitudinal dependence of the initial state to be a plane wave of momentum $p^\LCp$.  As the background has no longitudinal position dependence, longitudinal momentum $p^\LCp$ is conserved (both classically and quantum mechanically). Thus we need only consider the physics of the transverse directions. For the transverse degrees of freedom, we take the initial wavefunction to be a normalised Gaussian wave packet, also proportional to the HO eigenstate $\Phi^b_{00}(x^\LCperp)$. We will take the initial state width $b_{i}$ to be equal to electron mass $m_{e}$ because it is the only mass scale in QED. Hence the electron is localised to within $\sim m_{e}$ in transverse momentum space and $\sim 1/m_{e}$ in transverse position space. 

We evolve the wavefunction in time numerically by solving the Schr\"odinger equation Eq.~(\ref{eqn:time}) using the MSD2 scheme described above. This gives the wavefunction at each time step. In our calculation, the width of the 2D-HO basis in the transverse plane $b$ is chosen to coincide with the width of the initial state $b_{i}=m_{e}$, in order to improve convergence. The mod-square of the wavefunction then gives the particle probability distribution in either coordinate or momentum space, as desired. (See~\cite{Leutwyler:1977vy} for a comprehensive discussion of the position operator in light-front quantisation, and~\cite{Ilderton:2013dba} for an application to radiation reaction in the front form.)

\begin{figure}[t!]
   \centering
	\includegraphics[width=.5\textwidth]{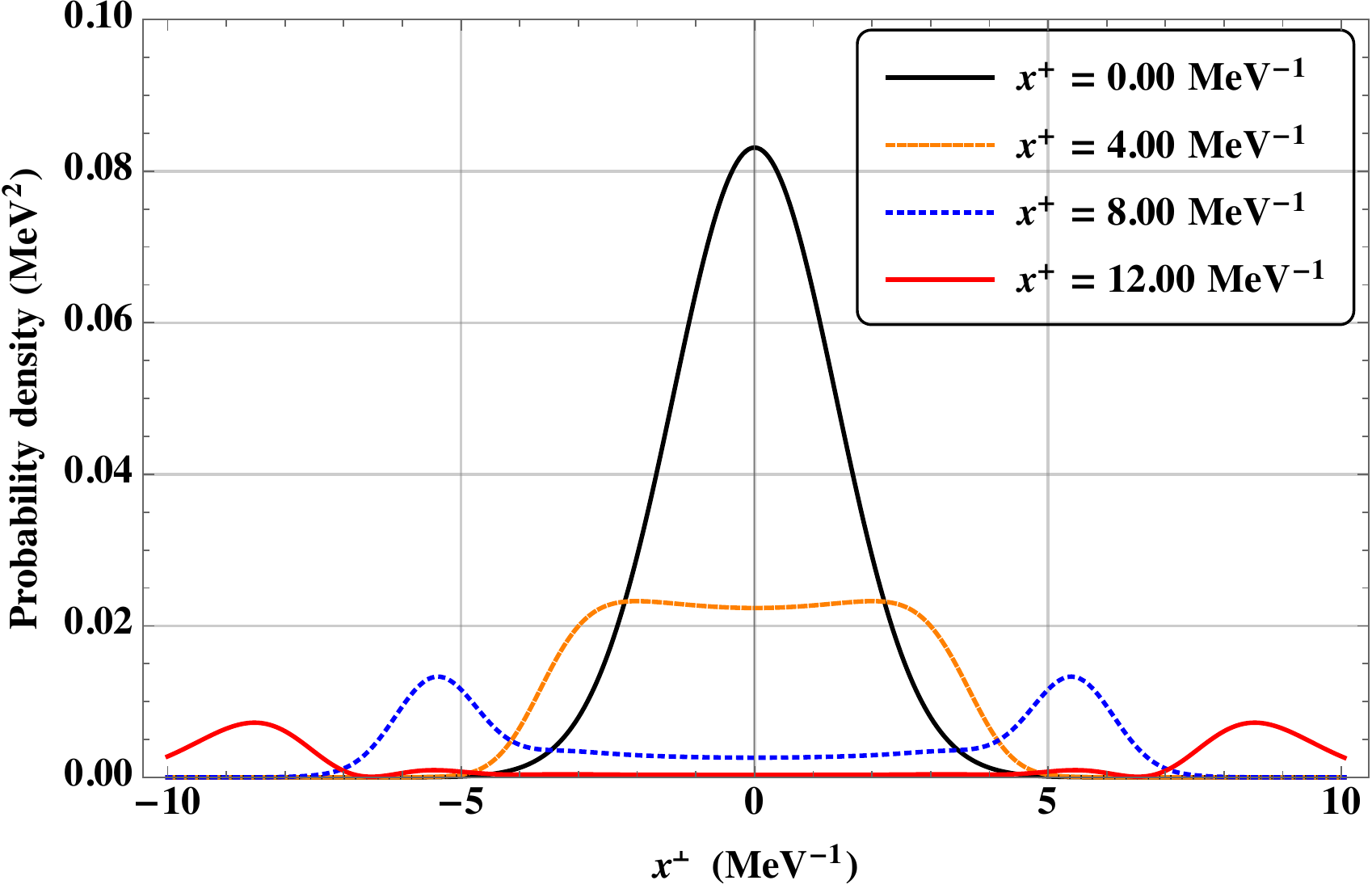}
      \caption{\label{fig:CMmotion}
      A single electron initially in a HO basis state is evolved through time in the axicon background with $a_0>0$. The wave packet is dispersed from the origin, effectively feeling a repulsive electromagnetic field.  Parameters: $N_{\rm max}=200$, $K_{\rm total}=10.5$, $b=m_{e}$, $b_{l}=m_{e}$, $a_{0}=10$, $b_{i}=m_{e}$, $p^{\LCp}=10.5\rm MeV$.}
\end{figure}

Fig.~\ref{fig:CMmotion} shows the time-evolution of the wave packet for $a_0>0$. (The background field Hamiltonian conserves angular momentum $L_z$, and our initial state is rotationally symmetric; hence it is enough to pick a slice through the transverse plane.) As time passes, the wave packet is dispersed and repelled from the origin, with the initial peak of high probability density being pushed to large transverse positions.  This is the analogue of the classical repulsion described above. We can also make a more quantitative check against the classical theory, as follows. The classical equations of motion reduce, in the transverse directions, to the coupled ODES
\be
	{\ddot x}^\LCperp = \frac{m_e a_0 b_l^2}{2p^\LCp\sqrt{\pi}}\; x^\LCperp \exp\bigg(-\frac{1}{2}b_l^2 x^\LCperp x^\LCperp\bigg) \;,
\ee
where a dot indicates an $x^\LCp$ derivative. We numerically integrate this equation for a large number of initial conditions sampled from the same Gaussian position and momentum space distributions as define the wave packet in the quantum calculation. The idea is then to track the position of the peak of the time-evolved distribution, expecting that this would be reproduced to some approximation by the peak of the quantum wave packet, where the probability density is highest. We find that the peak of the distribution shifts, in position space, out to $|x^\LCperp|  \approx 5.5$~MeV$^{-1}$  at $x^\LCp = 8$~MeV$^{-1}$ and to $| x^\LCperp|  \approx 8$~MeV$^{-1}$  at $x^\LCp = 12$~MeV$^{-1}$, in good agreement with the peak positions in the tBLFQ calculation; see Fig.~\ref{fig:CMmotion}. This is a very simple check, but it gives us confidence that we are capturing the correct physics both qualitatively and quantitatively. 

\begin{figure*}[htp]
   \centering
   \begin{center}
      \begin{tabular}{@{}cccc@{}}
         \includegraphics[width=.47\textwidth]{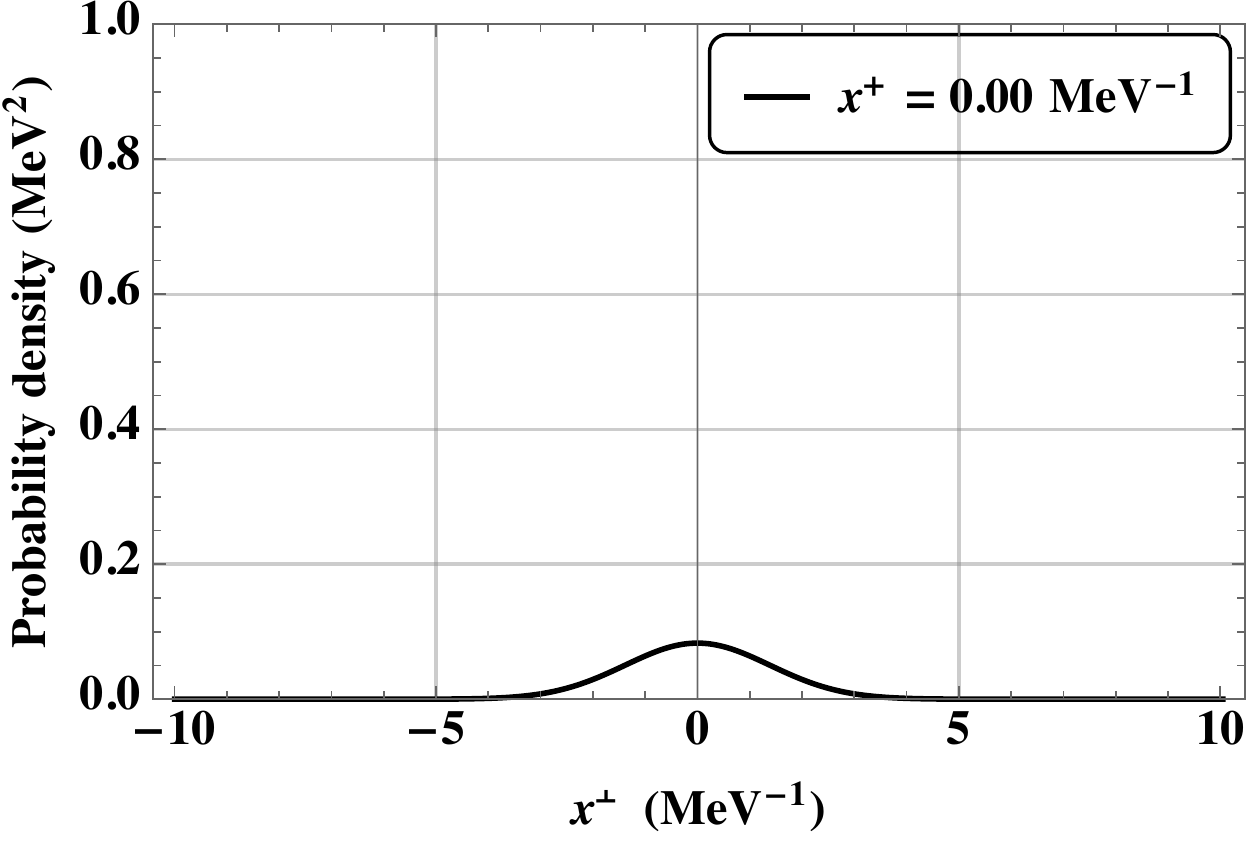} &
         \includegraphics[width=.47\textwidth]{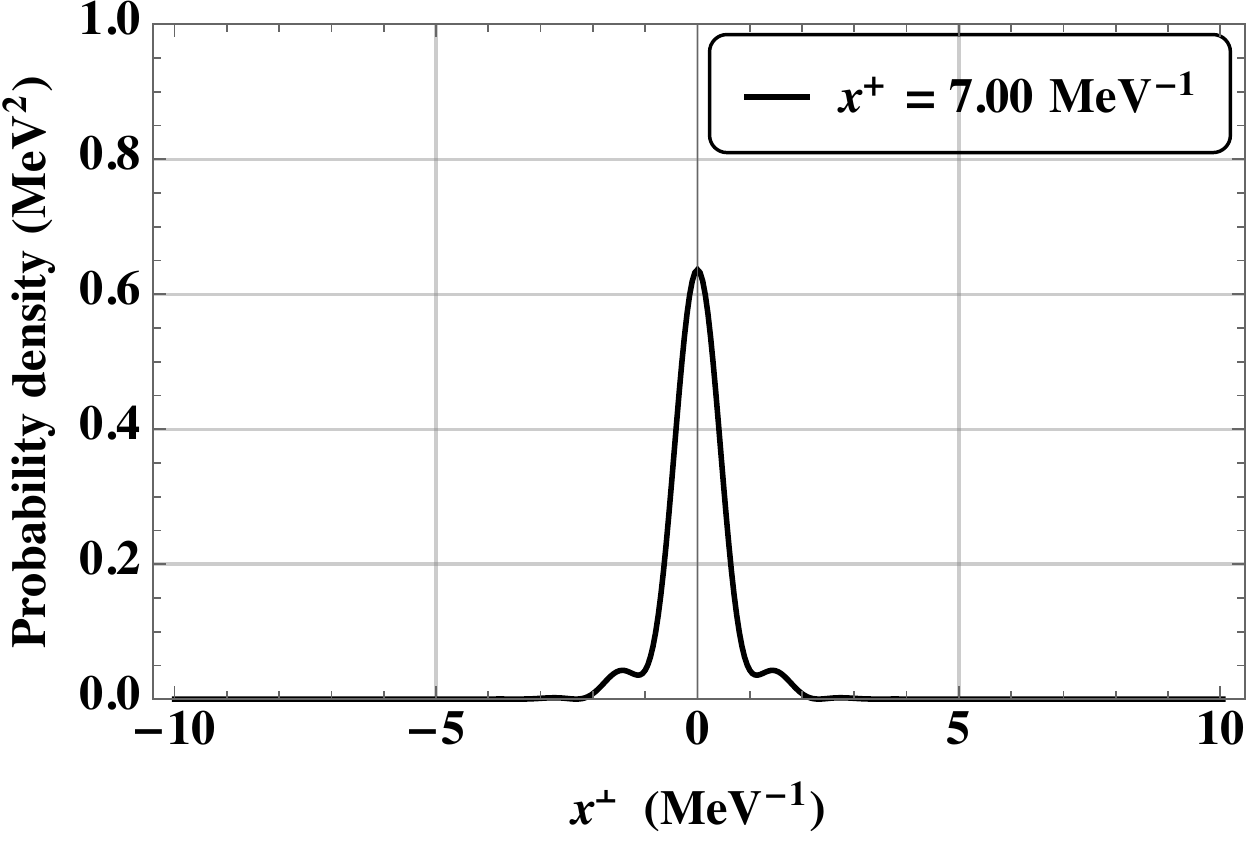} &\\
         \includegraphics[width=.47\textwidth]{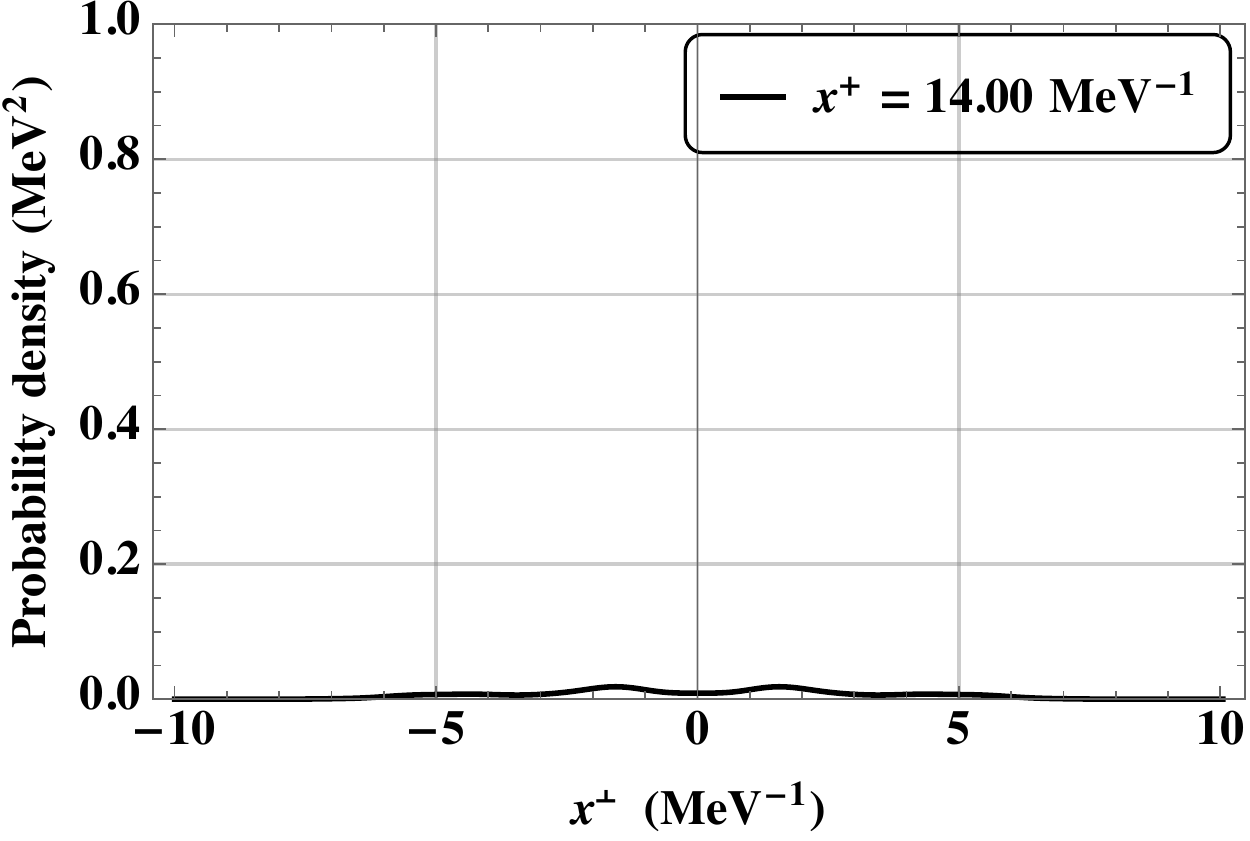} &
         \includegraphics[width=.47\textwidth]{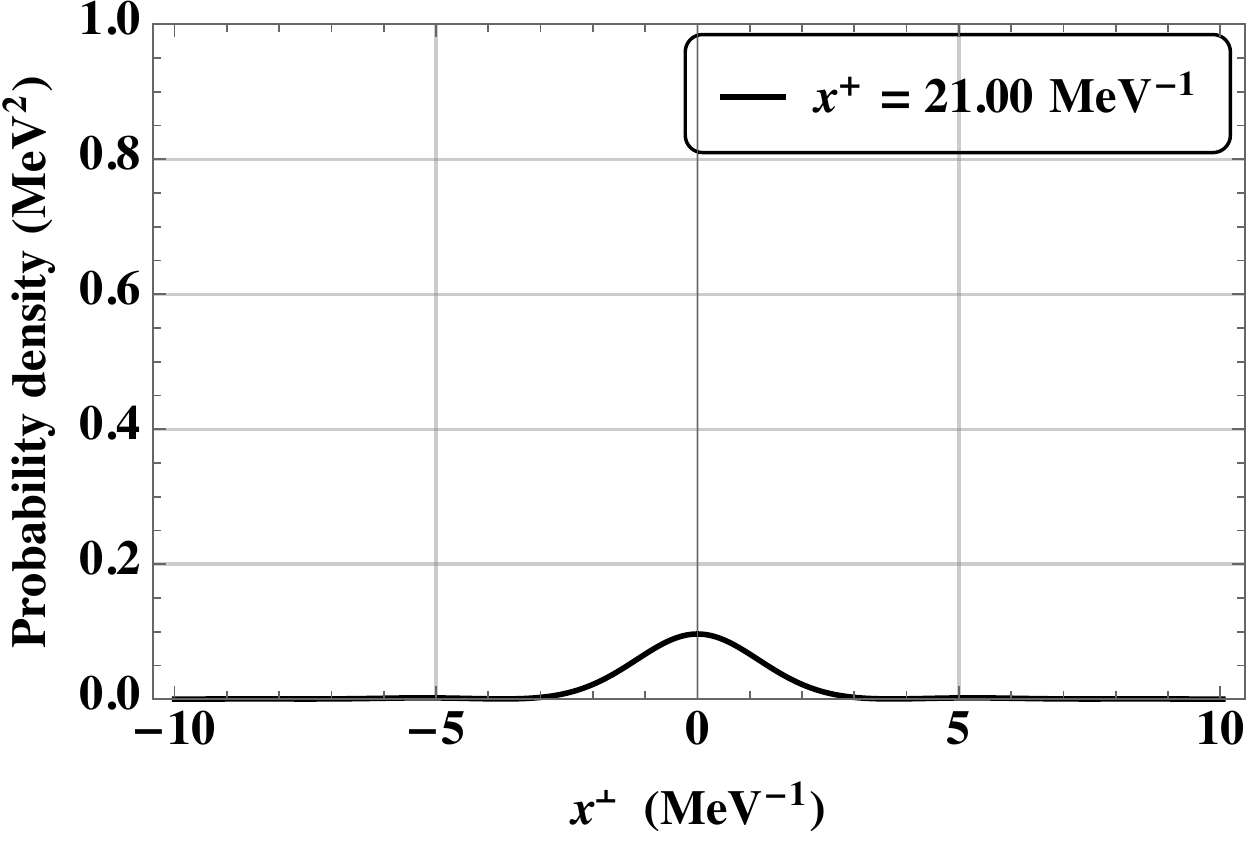} &\\
      \end{tabular}
      \caption{
      \label{fig:attractCMmotion}
   For $a_0<0$, the same initial state as in Fig.~\ref{fig:CMmotion} is first attracted to, and then shows oscillations around, the origin. Parameters: $N_{\rm max}=200$, $K_{\rm total}=10.5$, $b=m_{e}$, $b_{l}=m_{e}$, $a_{0}=-10$, $b_{i}=m_{e}$, $p^{\LCp}=10.5\rm MeV$. }
   \end{center}
\end{figure*}

In the quantum theory there is of course wave packet spreading even without an external field. We therefore continue to the case of $a_{0}<0$, all other parameters constant. This amounts to changing the sign of the electromagnetic field or, in this system, exchanging particles for antiparticles. The results are shown as a series of snapshots in Fig.~\ref{fig:attractCMmotion}.  In this case the initial wave packet is first compressed to the origin, with the probability density increasing there, before expanding back out, and being re-compressed toward the origin. This competition between wave packet focussing and spreading~\cite{Heinzl:2017zsr} naturally reflects the expected classical dynamics. The behaviour initially appears reminiscent of that of a harmonic oscillator in a potential trap. However, the Gaussian falloff of our background means that there is no true trapping, and for large times the probability density around the origin falls to zero.

%
\section{Photon emission}\label{SECT:PHOTONS}
%
In the Section above we discussed the time-evolution of a single electron in a background field, including only the $|\,e\,\rangle$ Fock state in the tBLFQ calculations. This represents the centre-of-mass (CM) motion (in that it resembles the CM motion in non-relativistic mechanics, as will be explained below). In this section, we add the second simplest Fock sector $|\,e\gamma\,\rangle$, with which we can study the photon emission and absorption. We are mainly interested in the production (stimulated by the background) of a physical photon from a physical electron, and the relative motion between them.  However, the CM spectrum is continuous and typically mixed with the relative motion, making the full energy spectrum difficult to handle, even in a discrete basis. Fortunately, in BLFQ, it is possible to factorise the CM and the relative motion. By adopting the `Lagrange multiplier' method \cite{Gloeckner:1974sst,lipkin1958hj}, we can isolate and study only the relative excitation.

As discussed in Sec.~\ref{SECT:BACKGROUND}, time-evolution is naturally studied in the tBLFQ basis $|\,\beta\,\rangle$ of QED eigenstates defined in Eq.~(\ref{eigenequation}). Calculations would be made even simpler if this basis had the property of factorisation, such that the CM and relative motion could be separated, for then it would be possible to perform the calculation in the space of `relative' variables. In the absence of a background (as in BLFQ), the wavefunction can indeed be factorised -- the problem we must confront is that this may not be true in a background field (as in tBLFQ). Indeed the CM motion is mixed with the relative motion, in which we are mainly interested. 

In what follows, we will in Sect.~\ref{SECT:FACTORIZATION} and \ref{SECT:WITHBACKGROUND} describe the factorisation in BLFQ, and then argue that we can \textit{approximately} evolve (using tBLFQ) the system inside its CM frame, for simplicity. In Sect.~\ref{SECT:OBSERVABLE} we define some useful observables which encode the physics of interest before going on to study their time dependence in Sect.~\ref{SECT:RESULT}.

\subsection{Factorisation in the tBLFQ basis}\label{SECT:FACTORIZATION}
%
In the light-front framework, the transverse boosts have the same algebra as the Galilean boosts, which implies that the CM motion can be decoupled from the relative motion as in non-relativistic many-body systems~\cite{Heinzl:2000ht}, with the longitudinal momentum fraction ${\mathrm x_{i}}=p^{\LCp}_{i}/P^{\LCp}$ playing the role of the non-relativistic mass. Thus the CM position is defined as $R^{\LCperp}=\sum_{i}{\mathrm x_{i}}x^{\LCperp}_{i}$, where $x^{\LCperp}_{i}$ is the single-particle coordinate. These features allow us to factorise the CM motion and the relative motion as in non-relativistic quantum mechanics. 

The QED Hamiltonian $P^{\LCm}_{\rm QED}$ (not including terms related to the background field) commutes with the CM (total) momentum $P^{\LCperp}$ as well as the CM Hamiltonian $(P^{\LCperp})^2/P^{\LCp}$. They therefore have simultaneous eigenstates, and the corresponding wavefunction can be factorised:
\begin{equation}
\begin{split}
   \text{in the $|\,e\,\rangle$ sector;} \qquad &\Psi(p_{1}^{\LCperp},k_{1},s_{1})=\langle\, p_{1}^{\LCperp},k_{1},s_{1}\,|\,\beta\,\rangle=\phi_{\rm CM}(p_{1}^{\LCperp})\psi_{\rm rel}(k_{1},s_{1})\;,\\
   \text{in the $|\,e\gamma\,\rangle$ sector;} \qquad & \Psi(p_{1}^{\LCperp},p_{2}^{\LCperp},k_{1},k_{2},s_{1},s_{2})=\langle\, p_{1}^{\LCperp},p_{2}^{\LCperp},k_{1},k_{2},s_{1},s_{2}\,|\,\beta\,\rangle=\phi_{\rm CM}(P^{\LCperp})\psi_{\rm rel}(q^{\LCperp},k_{1},k_{2},s_{1},s_{2})\;,
   \label{eqn:exfactor}
\end{split}
\end{equation}
where 1 (2) represents the electron (photon); $P^{\LCperp}=p_{1}^{\LCperp}+p_{2}^{\LCperp}$ is the CM momentum and $q^{\LCperp}=\mathrm x_{2}p_{1}^{\LCperp}-\mathrm x_{1}p_{2}^{\LCperp}$ is the relative momentum in the transverse directions.

In general, this factorisation will not be preserved when a finite (truncated) basis is used. There is however a method by which to retain exact factorisation in finite truncation~\cite{Li:2013cga}, as follows. We introduce a `modified CM Hamiltonian' $H_{\rm CM}$ defined by~\cite{Li:2013cga,Maris:2013qma} 
\begin{equation}
H_{\rm CM}=\left(\sum_i p^{\LCperp}_{i}\right)^{2}+b^{4}\left(\sum_{i}\mathrm{x}_ix^{\LCperp}_{i}\right)^{2}=(P^{\LCperp})^{2}+b^{4}(R^{\LCperp})^{2}\;.
   \label{}
\end{equation}
We then subtract from the QED Hamiltonian $P^{\LCm}_{\rm QED}$ the real CM motion $(P^{\LCperp})^{2}/P^{\LCp}$, and replace it with $H_{\rm  CM}$, as so:
\begin{equation}
   P^{\LCm}_{\rm QED}\to P^{\LCm}_{\rm QED}-\frac{(P^\LCperp)^2}{P^{\LCp}} + \frac{\lambda}{P^{\LCp}}\left(H_{\rm CM}-2b^{2}\right) \;,
   \label{eqn:modifiedcmhami}
\end{equation}
in which the term $-2b^2$ subtracts the zero-point energy, and $\lambda>0$ is a Lagrange multiplier~\cite{Gloeckner:1974sst,lipkin1958hj}. Unlike the original CM Hamiltonian, the modified $H_\text{CM}$ commutes with $P^{-}_{\rm QED}$ in finite \textit{truncated} BLFQ bases. As a result, exact factorisation of the wavefunction is achieved, despite the finite truncation~\cite{Li:2013cga}. Furthermore, by modifying the CM Hamiltonian as above, we lift the \textit{nonzero} CM excitations by $2\lambda(2N+|\,M|\,)b^{2}/P^{\LCp}$ \textit{without} altering the part of the spectrum with $N=M=0$, i.e.~the part without CM excitation. By choosing a sufficiently large $\lambda$, we lift all the CM excitations beyond the highest relative excitation of interest and thus we obtain the desired excitation spectrum of the relative motion.

Now, the modified CM Hamiltonian commutes with the full Hamiltonian; hence their common eigenstates and the CM motion have definite $N$ and $M$. The two-particle wavefunction can thus be factorised as
   \begin{equation}
      \begin{split}
            \Psi(p_{1}^{\LCperp},p_{2}^{\LCperp},k_{1},k_{2},s_{1},s_{2})&=\tilde\Phi_{NM}^{b}(P^{\LCperp})\psi_{\rm rel}(q^{\LCperp},k_{1},k_{2},s_{1},s_{2})\;,
               \label{eqn:factorizated}
               \end{split}
         \end{equation}
where $\tilde\Phi_{NM}^{b}$ is the CM motion wavefunction and a HO eigenstate (see Appendix~\ref{SECT:HOBASIS}) with $b=m_{e}=0.511\rm MeV$; $\psi_{\rm rel}$ is the relative wavefunction, which we will ultimately use to construct observables. This relative wavefunction can be expanded in the HO basis as
      \begin{equation}
        \psi_{\rm rel}(q^{\LCperp},k_{1},k_{2},s_{1},s_{2})=\sum_{nm} f_{nmk_{1}k_{2}s_{1}s_{2}}\tilde\Phi^{b_{b}}_{nm}(q^{\LCperp})\;,
        \label{eqn:rel_wf}
  \end{equation}
  where the width $b_{b}:=\sqrt{\mathrm x_{1}\mathrm x_{2}}b$, and an $\mathrm x$ dependence is introduced for the exact factorisation in finite truncated bases~\cite{Li:2013cga}. The ${\mathrm x}_j$ are the longitudinal momentum fractions of the electron, $\mathrm x_{1}=k_{1}/(k_{1}+k_{2})$, and the photon, $\mathrm x_{2}=1-\mathrm x_{1}$. The coefficients $f$ can be obtained by the Talmi-Moshinksy transformation~\cite{talmi1952nuclear,Moshinsky:1959qbh,Li:2013cga}, which is typically used to separate the CM motion and the relative motion in the HO basis.
%
%
%
\subsection{Relative motion in a background field}\label{SECT:WITHBACKGROUND}
%
%
%
%
In the presence of a background field, the full Hamiltonian $P^{\LCm}(x^{\LCp})=P^{\LCm}_{\mathrm{QED}}+V(x^{\LCp})$ may not commute with the CM Hamiltonian $(P^{\LCperp})^{2}/P^{\LCp}$. In this case there is no longer an exact factorisation of the wavefunction as in Eq.~(\ref{eqn:exfactor}). This means that the relative motion will be affected by the CM motion and thus cannot be studied separately. 

We saw in Sec.~\ref{SECT:SINGLEELECTRON}, however, that the CM motion strongly resembles the classical motion. We thus make the approximation that the CM motion can be replaced with that of a classical electron in the background. The system then has a time-dependent CM momentum, and at any given time we can always boost the system to its CM frame. In this frame, there is no CM motion, while the background field changes by the \textit{inverse} boost.  Now, because the background (\ref{BG-AI}) has only a longitudinal component, it is left invariant under a boost in the transverse directions,
   \begin{equation}
      \begin{split}
         \mathcal{A}^{\LCp} \rightarrow \mathcal{A}^{\LCp}\;, \qquad\qquad
         {\mathcal{A}}^{\LCperp} \rightarrow {\mathcal{A}^{\LCperp}}+\mathcal{A}^{\LCp} {C}^{\LCperp}\;, \qquad\qquad
         \mathcal{A}^{\LCm} \rightarrow \mathcal{A}^{\LCm}+\mathcal{A}^{\LCperp}\cdot C^{\LCperp}+\mathcal{A}^{\LCp}C^{\LCperp2}\;, 
      \end{split}
   \end{equation}
   where the $C^{\LCperp}$ are dimensionless numbers that depend on the boost. What we should consider is the change of field caused by the displacement of the field from the origin in the CM frame. For now, we neglect this displacement of the background field. This approximation is valid when the single-electron wavefunction is initially centred on the origin where the electric field is small. We can then simply evolve the system inside the subspace with no CM motion, that is the subspace $N=M=0$, with the original background field in Eq.~(\ref{eqn:backgroundfield}). This is equivalent to saying that we only evolve the relative wavefunction $\psi_{\rm rel}$ in Eq.~(\ref{eqn:factorizated}).

%
\subsection{Observables}\label{SECT:OBSERVABLE}
%
We retain, from the whole Fock space, only the $|\,e\,\rangle$ and $|\,e\gamma\,\rangle$ Fock sectors. In the latter, we can study photon emission and the relative motion between the electron and the emitted photon. However, the $|\,e\gamma\,\rangle$ sector also contains non-physical photons, namely those `bare' photons which contribute to the physical electron. These are not real photons that are emitted and can be measured by a detector. However, we find in our calculations that the excited states of $P^{\LCm}_{\rm QED}$ are dominated by the $|\,e\gamma\,\rangle$ sector; the overlap of excited states with the $|\,e\,\rangle$ sector, mod-squared, being less than $10^{-3}$ at $N_{\rm max}=100, K_{\rm total}=10$. This implies that the electron and the photon are weakly coupled, and therefore we approximate the excited states (of $P^{\LCm}_{\rm QED}$) as scattering states of a physical electron and a physical photon. 
In this paper, we focus on the physical part of the wavefunction $\bar \Psi$ which is obtained by projecting out the physical electron state
\begin{equation}
   \bar\Psi(p_{1}^{\LCperp},p_{2}^{\LCperp},k_{1},k_{2},s_{1},s_{2};x^{\LCp}):=\langle\, p_{1}^{\LCperp},p_{2}^{\LCperp},k_{1},k_{2},s_{1},s_{2}|\,(\mathrm I-|\,e\,\rangle_{\rm phys} \langle\, e\,|_{\rm phys})|\,\Psi;x^{\LCp}\,\rangle_{\rm S}\;,
   \label{eqn:phys_wf}
\end{equation}
where $|\,e\,\rangle_{\rm phys}$ is the physical electron state; $|\,\Psi;x^{\LCp}\,\rangle_{\rm S}$ is the Schr\"odinger picture state vector obtained by using Eq.~(\ref{eqn:time}) and~(\ref{eqn:interactionpictureoperator}). Physically, this eliminates the non-physical bare photon contribution. It is easiest to achieve the projection in the $|\,\beta\,\rangle$ basis, where it corresponds to simply dropping the lowest component of the wavefunction.

It would be desirable to obtain the scattering matrix in tBLFQ; however, the systematic construction of `out states'~\cite{Weinberg:1995mt} is a topic for future work. In this paper we instead concentrate on two observables at finite time. They are the longitudinal momentum distribution (LMD), which describes the longitudinal motion, and the transverse momentum distribution (TMD), which describes the transverse motion.\footnote{Since the momentum distribution is more commonly measured in experiments, we will focus on the distribution in momentum space in the remainder of the paper.} These are defined in terms of the \textit{physical} part of the relative wavefunction as
\begin{eqnarray}
  \label{pdf}
  \mathrm {LMD} (\mathrm{x}_{1}; x^{\LCp}) &:=& \sum_{s_{1}s_{2}}\int d^{2}q^{\LCperp}\left|\,\bar\psi_{\rm rel}(q^{\LCperp},k_{1},k_{2},s_{1},s_{2}; x^{\LCp})\right|\,^{2}\;, \qquad \mathrm{x_{1}}=\frac{k_{1}}{k_{1}+k_{2}} \;, \\
  \label{TMD}
  \mathrm {TMD}\left(q^{\LCperp},\mathrm x_{1};x^{\LCp}\right) &:=& \sum_{s_{1}s_{2}}\left|\,\bar\psi_{\rm rel}(q^{\LCperp},k_{1},k_{2},s_{1},s_{2};x^{\LCp})\right|\,^{2}\;,
  \end{eqnarray}
in which the relative wavefunction $\bar\psi_{\rm rel}$ is obtained by a TM transformation from the two particle wavefunction (\ref{eqn:phys_wf}), as in Eq.~(\ref{eqn:factorizated}) and~(\ref{eqn:rel_wf}). The meaning of the LMD is that it gives the probability of finding a scattered electron-photon state with an electron longitudinal momentum fraction $\mathrm x_{1}=k_{1}/(k_{1}+k_{2})$.
The TMD tells us the probability of finding the scattered electron-photon states with relative \textit{transverse} momentum $q^{\LCperp}$, and momentum fraction of the physical electron, $\mathrm x_{1}$. If, in Eq.~(\ref{TMD}), we drop the sum over $s_{1}$ and $s_{2}$ then we obtain a spin-dependent TMD. If the TMD~(\ref{TMD}) is integrated over $q^{\LCperp}$ then we recover the LMD~(\ref{pdf}).

\subsection{Time-evolution in tBLFQ}\label{SECT:RESULT}
%
%
In this section we discuss the time-evolution of the LMD and TMD, as defined above. We start with the physical electron state, which is the lowest eigenstate of $P^{\LCm}_{\mathrm{QED}}$ \textit{without} the background field. Evolving this state in time, with the background~(\ref{eqn:backgroundfield}) present, we obtain the wavefunction at each time step. (This is performed numerically using the MSD2 scheme described in Sec.~\ref{SECT:SEQUATION}.) Equations~(\ref{pdf}) and~(\ref{TMD}) then  tell us how to extract the time-evolution of the LMD and the TMD, respectively. 

\begin{figure*}[]
\centering
   \begin{center}
  \begin{tabular}{@{}cccc@{}}
     \includegraphics[width=.47\textwidth]{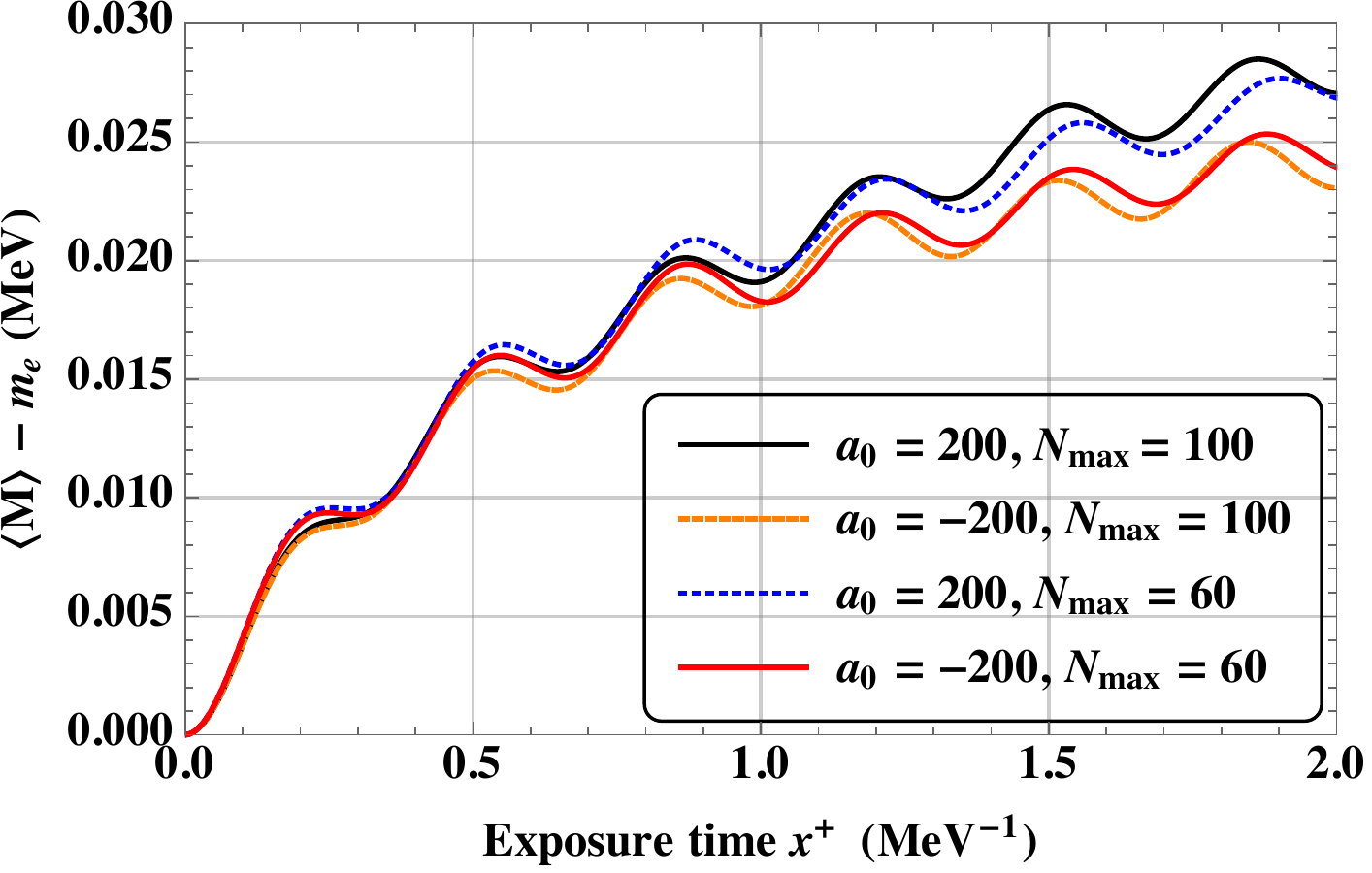} &
  \end{tabular}
  \caption{Probability to find electron-photon states, or rather the probability of photon emission. This probability initially grows with time. The later oscillations are discussed in the text. Parameters: $N_{\rm max}=100$ and $60$, $K_{\rm total}=10.5$, $b_l=m_{e}$, $a_0=\pm200$, $P^{\LCp}=10.5\rm MeV$.}
   \label{fig:photonprobability}
   \end{center}
\end{figure*}

In order not to over-complicate the presentation and obtain a broad overview of the physics involved, we start by considering a simple quantity, namely the probability to find a physical photon of \textit{any} (longitudinal or transverse) momentum; this is obtained by integrating the LMD over all longitudinal momentum fractions $\mathrm x_{1}$. Fig.~\ref{fig:photonprobability} shows this probability as a function of time in backgrounds with  $a_{0}=\pm200$, i.e. fields which give, broadly speaking, repulsive and attractive forces to the electron, respectively. Initially, at light-front time $x^\LCp=0$, there is only a physical electron, and no physical photons; hence the probability starts from  zero. As time passes the probability increases as electrons are raised to excited states by the background. Initially, the probability of photon emission is independent of the sign of $a_0$, but at later times the sign of the field begins to have an effect.  As can be seen in the figure, all the curves show a similar trend -- the probability of photon emission increases, but with small oscillations. This reflects the exchange of energy/momentum back and forth between the electrons and produced photons  as time passes. We observe that, numerically, these oscillations  come from the matrix element $\langle\, \rm lowest~state\,|\,V\,|\,\rm lowest~state\,\rangle$, which is significantly bigger than the other matrix elements. 

We now turn to what are essentially the \textit{differential} photon emission probabilities. We begin by studying, through the TMD, the time-evolution of only the \textit{transverse} motion of the physical electron, fixing $k_{1}=5.5$, i.e.~the \textit{longitudinal} momentum fraction $\mathrm x_{1}=5.5/10.5\approx0.52$. We will study how this depends on $N_{\rm max}$ and $a_{0}$. Finally, we will reinstate the longitudinal dependence, allowing us to study the time-evolution of the TMDs with different longitudinal momentum fractions $\mathrm x_{1}$ and the LMD. 

%
\subsubsection{$N_{\rm max}$ dependence}
%
\begin{figure*}[t!]
\centering
   \begin{center}
  \begin{tabular}{@{}cccc@{}}
     \includegraphics[width=.49\textwidth]{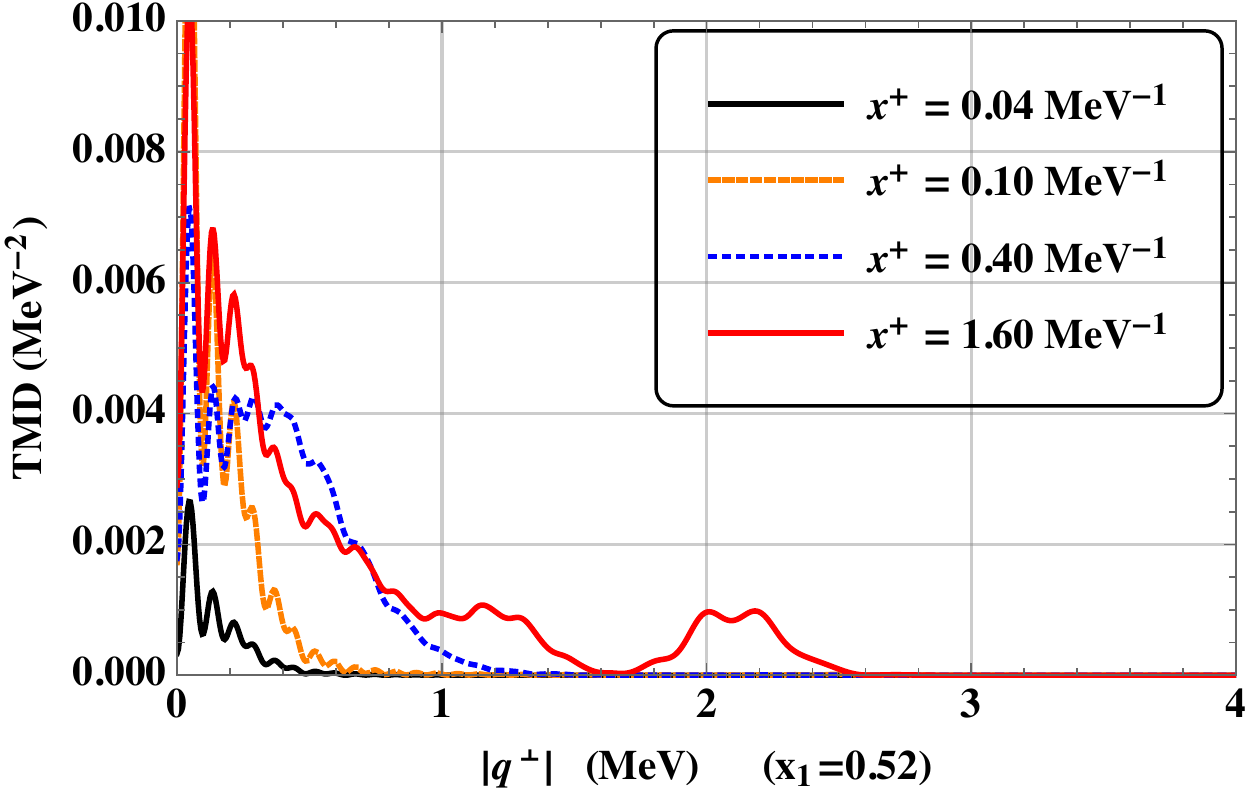} &
     \includegraphics[width=.49\textwidth]{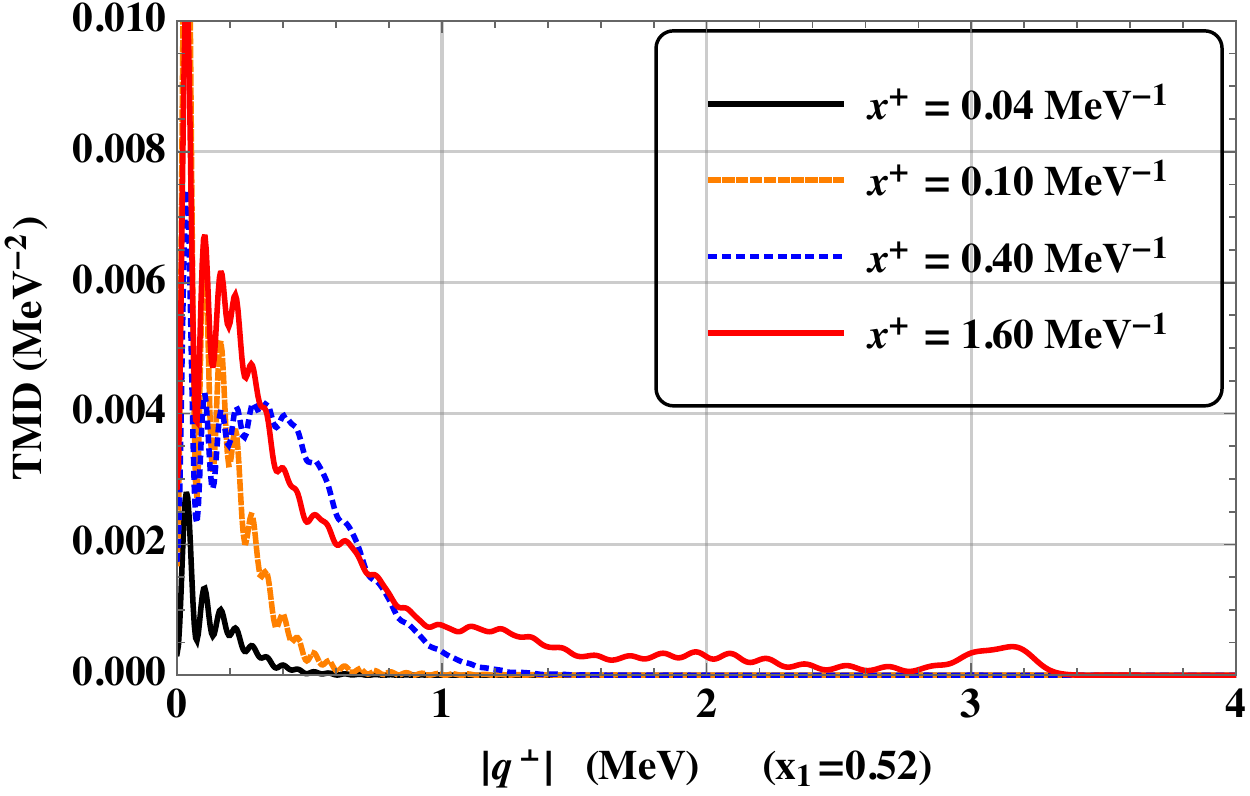}
  \end{tabular}
  \caption{Time-evolution of the TMD in two bases with (left) $N_{\rm max}=60$ and (right) $N_{\rm max}=100$. All other parameters are the same in both calculations: $K_{\rm total}=10.5$, $b_{ b}=0.50m_{e}$, $b_l=m_{e}$, $a_0=200$, $\mathrm x_{1}=0.52$, $P^{\LCp}=10.5\rm MeV$. Initially there is only a physical electron. As time passes, the transverse momenta are excited to higher values. For the smaller basis (left-hand panel), the excitations hit the $\rm UV$ cutoff (at around $2.6 \rm MeV$) at a time $x^{\LCp}\approx  1.2 \rm MeV^{-1}$, whereas for the larger basis (right-hand panel), the UV cutoff (now at $3.4 \rm MeV$ due to the larger $N_{\rm max}$) at a later time of $x^{\LCp}\approx 1.6 \rm MeV^{-1}$. Overall, there is little difference between the curves until they hit the UV cutoff. Beyond the corresponding `cutoff' times, the curves are no longer reliable.}

   \label{fig:nmaxdependence}
   \end{center}
\end{figure*}
We should of course address how our results depend on the various parameters defining our basis. Here we focus on $N_{\rm max}$, which is a truncation parameter for transverse degrees of freedom. It limits the energy (\ref{hoenergy}) of the 2D-HO basis. In a basis with a larger $N_{\rm max}$, we have a higher UV cutoff, proportional to $\sqrt{N_{\rm max}}$, and a lower IR cutoff, proportional to $1/\sqrt{N_{\rm max}}$ in momentum space \cite{Zhao:2013cma}.
Thus, larger $N_{\rm max}$ offers a more complete description of the system. However, Fig.~\ref{fig:photonprobability} shows that the time-evolution of the probability of photon emission only depends weakly on $N_{\rm max}$, at least initially.\footnote{In our calculation, we retain only the lowest $1921$ ($1121$) eigenstates of $P^{\LCm}_{\rm QED}$, which have no CM excitation, for $N_{\rm max}=100, K_{\rm total}=10$ ($N_{\rm max}=60, K_{\rm total}=10$); this means that we retain the states with invariant mass less than $6.67\rm MeV$ ($5.02\rm MeV$).}

Fig.~\ref{fig:nmaxdependence} shows the time-evolution of a TMD in the background~(\ref{eqn:backgroundfield}) with $a_{0}=200$. The two panels show the same calculation performed for two values of $N_{\rm max}$, namely 60 (left) and 100 (right). Now, while the background does not couple directly to the photon, it can excite the electron to higher excited states in which the relative motion between the electron and the photon is more significant; this is clear in the figures. As is shown by comparing the two panels, the overall shape of the TMD is not strongly dependent on $N_{\rm max}$, with the main difference being that for the larger $N_{\rm max}$ the curve is smoother, since the resolution is better. This is, however, all \textit{provided} that the momentum excitations do not reach the `boundary' implied by the UV cutoff of the basis, beyond which the results are subject to truncation artifacts. For the larger basis, the curves hit the (larger) cutoff at later times.  (We remark that while there exist `rescaling' methods~\cite{Zhao:2014hpa} which improve the convergence of the wavefunction destroyed by the Fock-sector truncation, we need not use the method in this paper because of the good convergence of our results with $N_{\rm max}$.)

\begin{figure*}[t!]
\centering
   \begin{center}
  \begin{tabular}{@{}cccc@{}}
     \includegraphics[width=.47\textwidth]{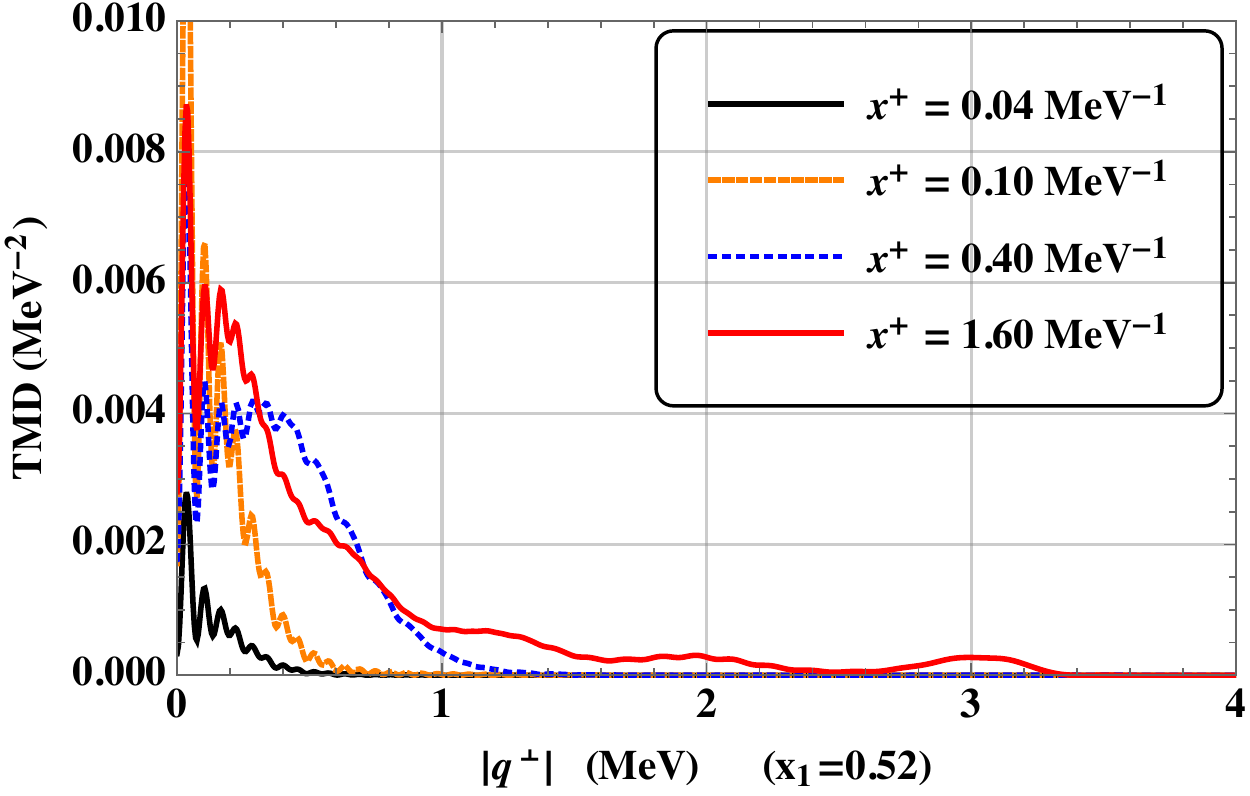} &
     \includegraphics[width=.47\textwidth]{./figures/relative/n100k10e100ks5_2.pdf} &\\
     \includegraphics[width=.47\textwidth]{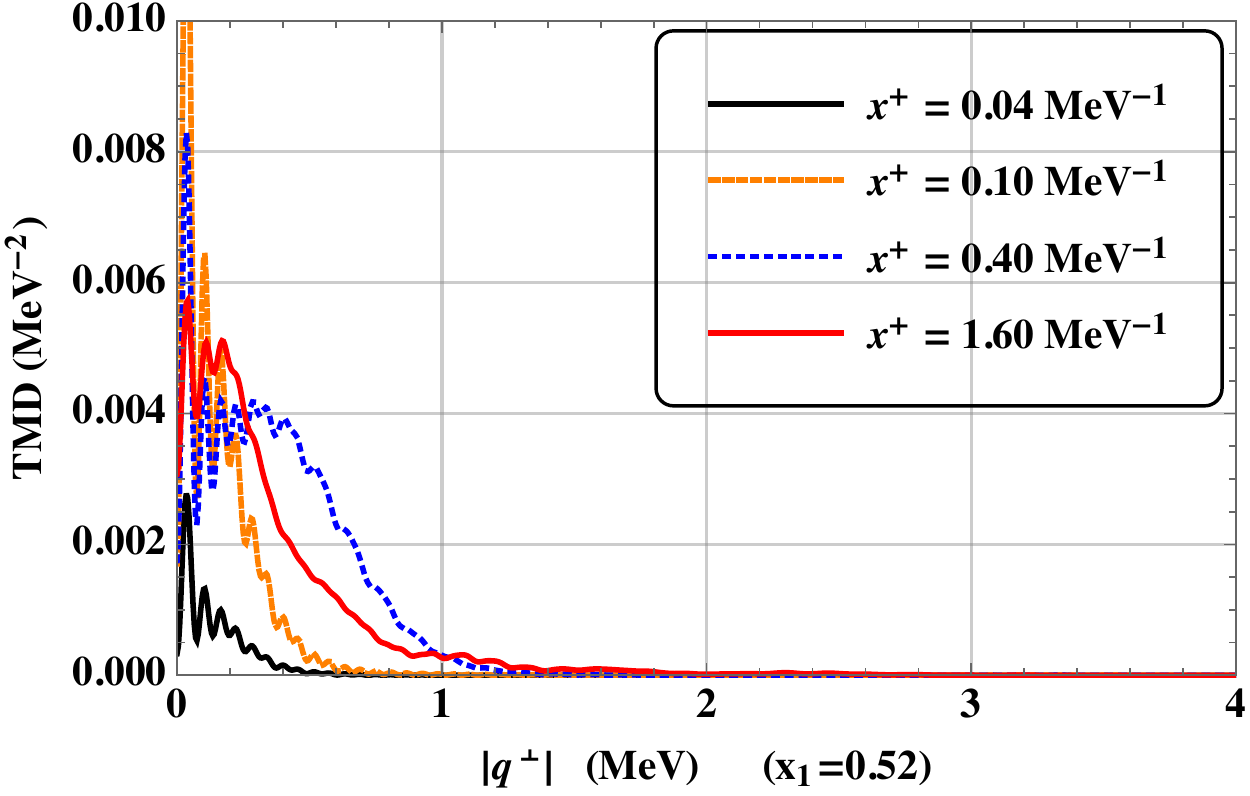} &
     \includegraphics[width=.47\textwidth]{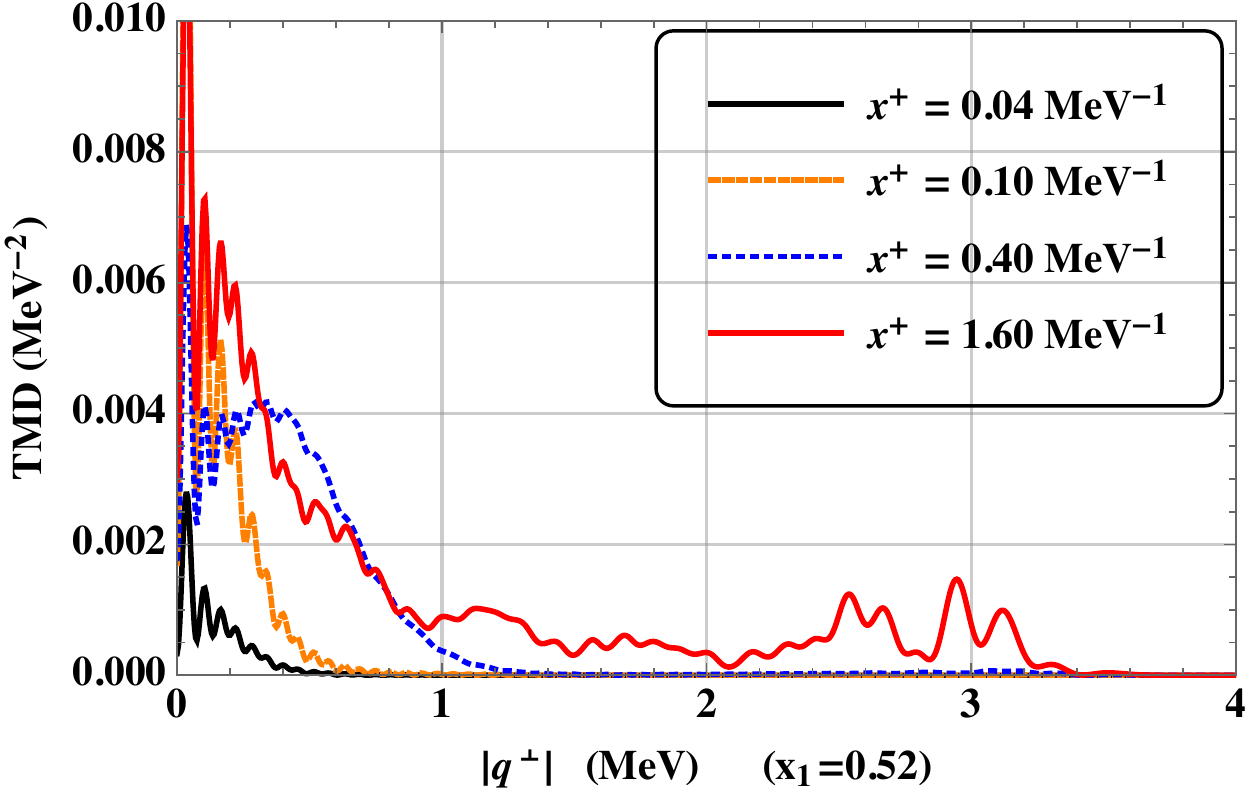} &
  \end{tabular}
  \caption{Time-evolution of the TMD. In all plots, $N_{\rm max}=100$, $K_{\rm total}=10.5$, $b_{ b}=0.50m_{e}$, $b_l=m_{e}$, $\mathrm x_{1}=0.52$. In the left-hand column, $a_0=-200$; in the right-hand column $a_0=200$. Longitudinal momentum $P^{\LCp}$ decreases from $P^{\LCp}=10.5\rm MeV$ to   $P^{\LCp}=1.05\rm MeV$ as we move from the upper to the lower row. In the left column (attractive field), as $P^{\LCp}$ decreases, the high momentum excitations decrease. Contrastingly, in the right column (repulsive field), the higher momentum excitations grow as $P^{\LCp}$ decreases.}
   \label{fig:signdependence}
   \end{center}
\end{figure*}
\subsubsection{$a_{0}$ dependence and $P^{\LCp}$ dependence}\label{SECT:ADEPENDENCE}

Following the above, we now fix $N_{\rm max} = 100$. In Fig.~\ref{fig:signdependence} we consider the dependence of the TMD on  the sign  of~$a_0$, i.e.~on whether the background is attractive or repulsive, and on the total longitudinal momentum of the physical electron $P^{\LCp}$. We begin with  $P^{\LCp}=10.5\rm MeV$ as shown in the upper two panels of the figure. In the left-hand panel, $a_0=-200$ and  the field is essentially attractive. In the right-hand panel, $a_0=+200$ and the field is repulsive. In both cases, as time passes, the TMDs become distributed across an increasingly broad range of momenta. The TMDs are almost identical for short times, while for longer larger times there are, by time $x^{\LCp}=1.60\rm MeV^{-1}$, slightly more high momentum excitations for $a_0>0$. (In both cases the curves hit the boundary at about $x^{\LCp}=1.60 \rm MeV^{-1}$ and are then subject to artifacts.)

Figs.~\ref{fig:photonprobability} and \ref{fig:signdependence}, as well as Figs.~\ref{fig:spinprobability} and \ref{fig:invmass} below, all show that the short-time evolution of the physical electron is not strongly dependent on the sign of the background, at least for $P^{\LCp}=10.5\rm MeV$.  We attribute this to the large light-front momentum chosen, implying high energy, for which accelerated charges emit in a narrow cone around their forward direction (instantaneous synchrotron radiation~\cite{Jackson:1998nia}); hence it makes little difference whether the background is attractive or repulsive.

In order to accentuate the differences due to the sign of $a_{0}$, we therefore reduce the total longitudinal momentum~$P^{\LCp}$ by changing the length of the circle in Eq.~(\ref{eqn:disc_longitudinal}) from\footnote{The choice of $L=2 \pi \mathrm{MeV}^{-1}$ allows the longitudinal quantum number $k$ to be interpreted as the longitudinal momentum in units of ${\rm MeV}$.} $L=2 \pi \mathrm{MeV}^{-1}$ to $L=20 \pi \mathrm{MeV}^{-1}$. As a result, $P^{\LCp}$ becomes equal to $0.1K_{\rm total}$ in units of $\rm MeV$. The resulting time-evolution of the electron with this smaller $P^{\LCp}$ is shown in the lower panels of Fig.~\ref{fig:signdependence} (see also Figs.~\ref{fig:spinprobability} and \ref{fig:invmass}, below). In this figure  $K_{\rm total}=10.5$ but the longitudinal momentum $P^{\LCp}$ equals $10.5\rm MeV$ or $1.05\rm MeV$. As expected, there are larger differences visible in the plots for opposite signs of $a_0$  when $P^{\LCp}$ is smaller. 

\subsubsection{Time-evolution in different spin components}

\begin{figure*}[t]
\centering
   \begin{center}
  \begin{tabular}{@{}cccc@{}}
     \includegraphics[width=.47\textwidth]{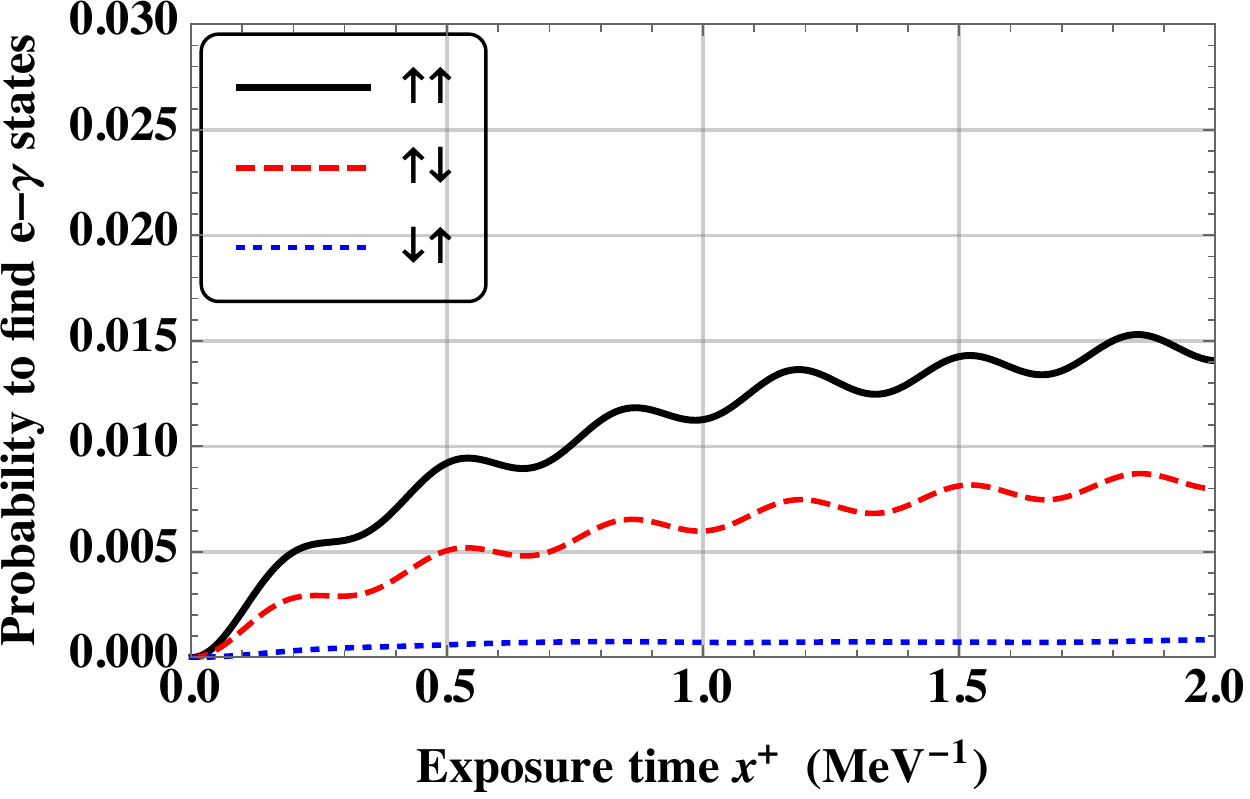} &
     \includegraphics[width=.47\textwidth]{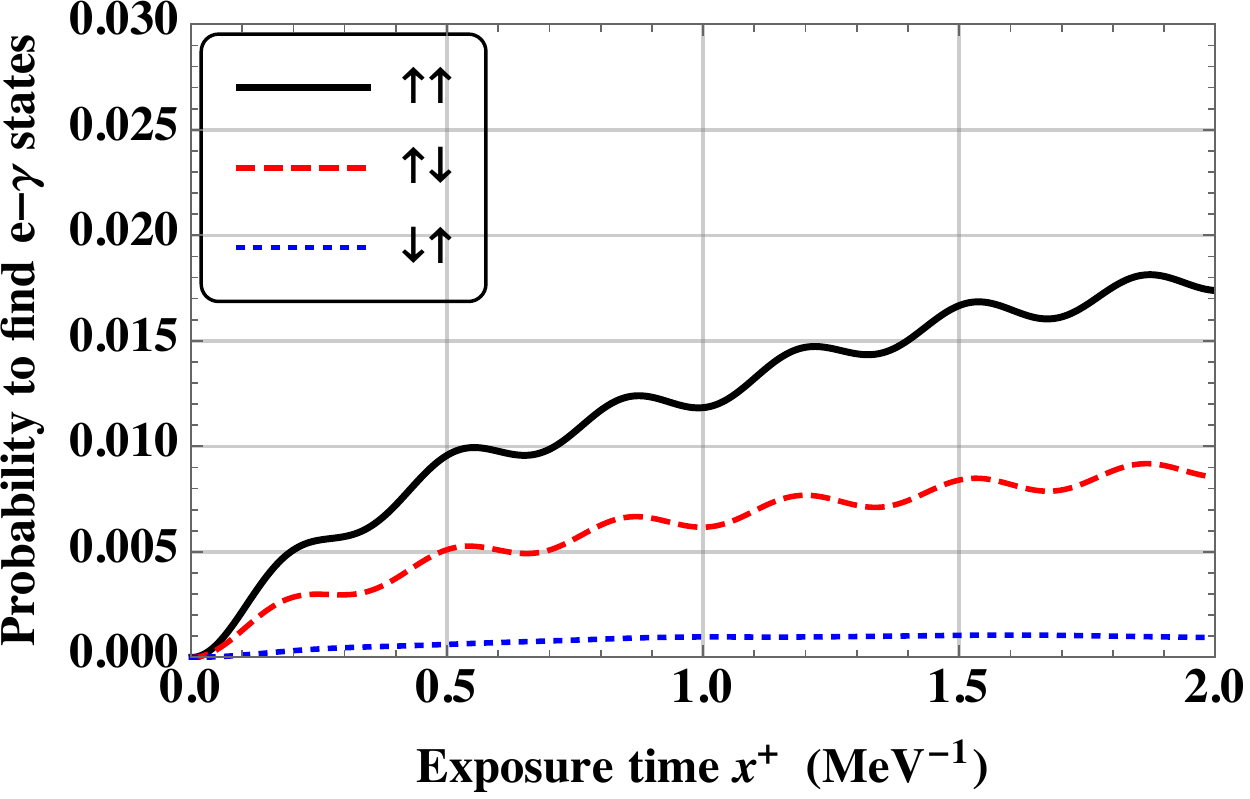} &\\
     \includegraphics[width=.47\textwidth]{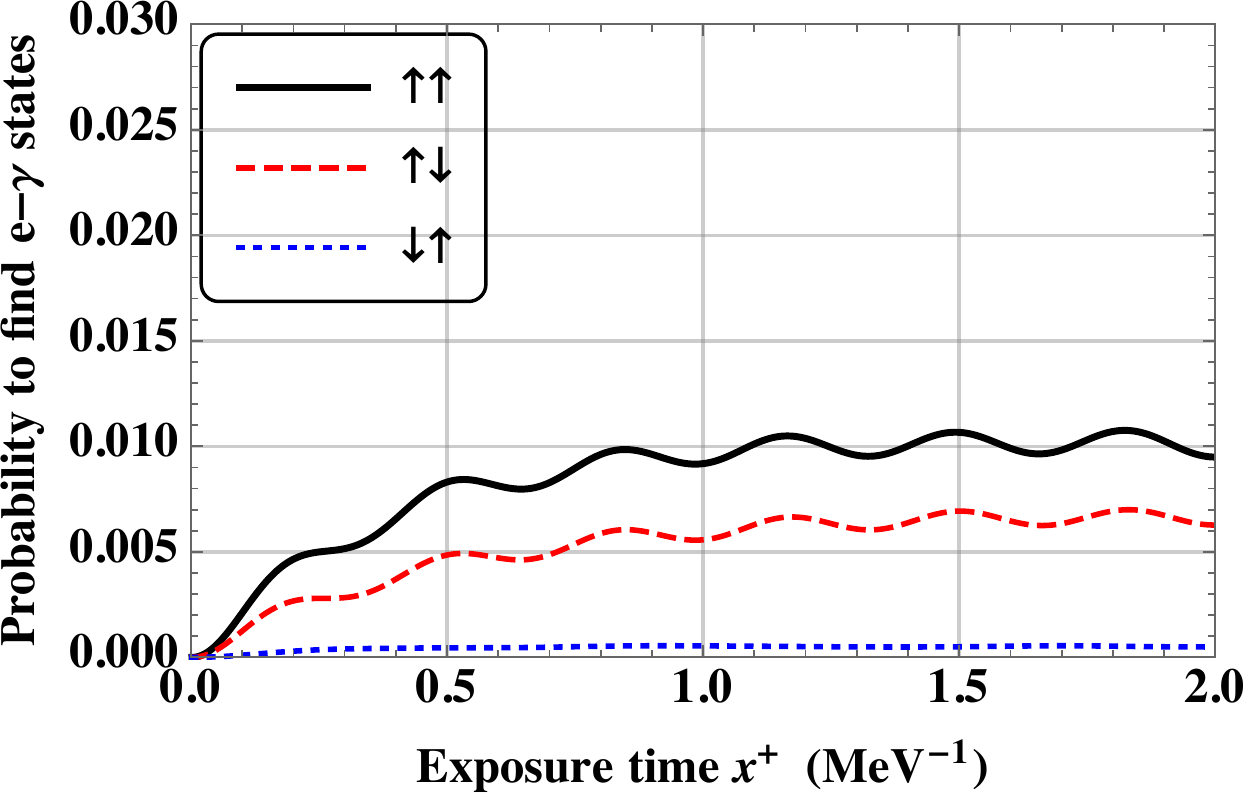} &
     \includegraphics[width=.47\textwidth]{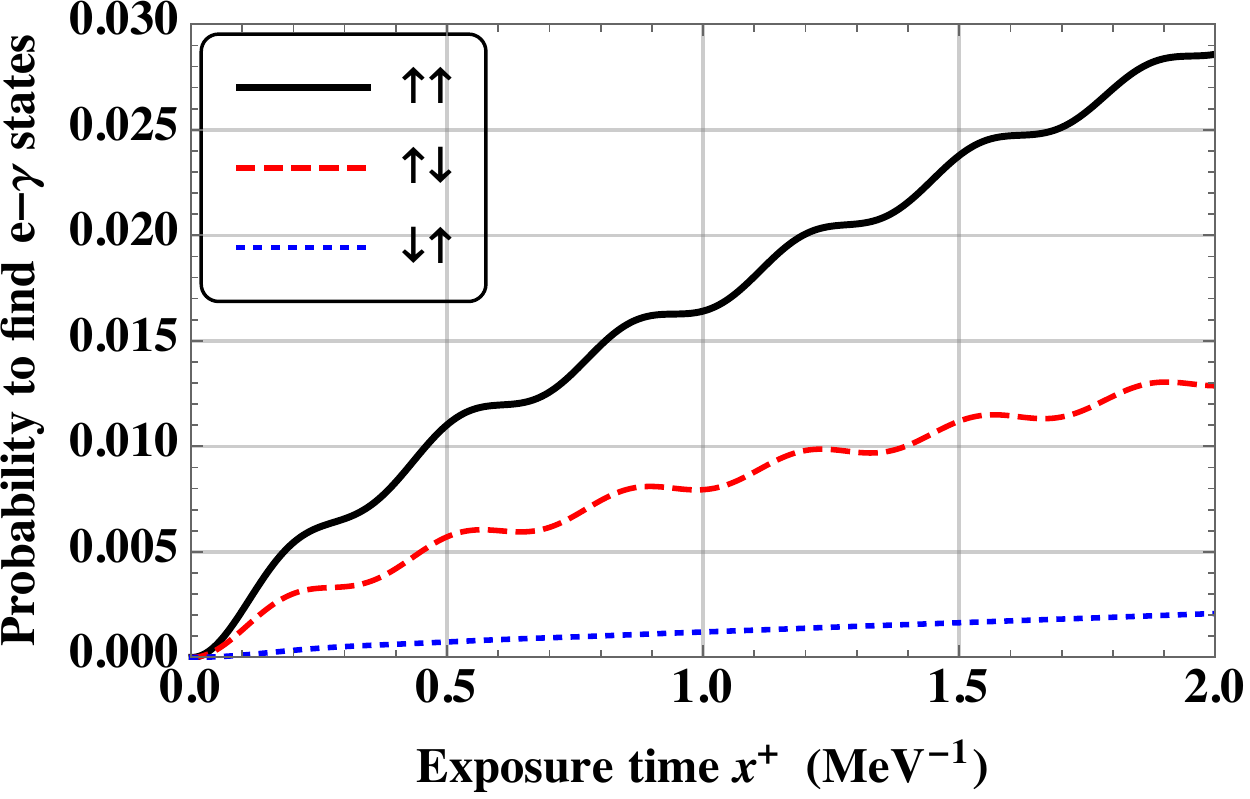} &
  \end{tabular}
  \caption{\label{fig:spinprobability}Spin-resolved probability of finding electron-photon states. In all plots, $N_{\rm max}=100$, $K_{\rm total}=10.5$, $b_l=m_{e}$. In the left-hand column, $a_0=-200$; in the right-hand column $a_0=200$. Longitudinal momentum decreases from $P^{\LCp}=10.5\rm MeV$ to  $P^{\LCp}=1.05\rm MeV$ as we move from the upper to the lower panels . At higher momentum, the probability of finding any particular spin configuration increases faster for $a_0>0$ than for $a_0<0$. At lower momentum, the curves show more differences between $a_{0}>0$ and $a_{0}<0$.}
   \end{center}
\end{figure*}

Fig.~\ref{fig:spinprobability} shows the spin-resolved probability of photon emission. We consider the electron/photon spin configurations $\uparrow\uparrow$, $\uparrow\downarrow$, $\downarrow\uparrow$. (The $\downarrow\downarrow$ contribution is zero and not included in the plots.) As seen in each of the four panels, the spin-flip component $\downarrow\uparrow$ is very small. The reason for this is that the background does not directly change the particle spin, and the spin-flip matrix elements from the background-free parts of the  QED Hamiltonian are small~\cite{Zhao:2014xaa}. Fig~\ref{fig:spinprobability} also shows that there is a more significant dependence on the sign of the field for smaller $P^\LCp$ than for larger, as was also observed above. Note that if we sum over all the spin configurations, we recover Fig.~\ref{fig:photonprobability} as we should.

%
\subsubsection{Invariant mass time-evolution}
\begin{figure*}[t]
\centering
   \begin{center}
  \begin{tabular}{@{}cccc@{}}
     \includegraphics[width=.5\textwidth]{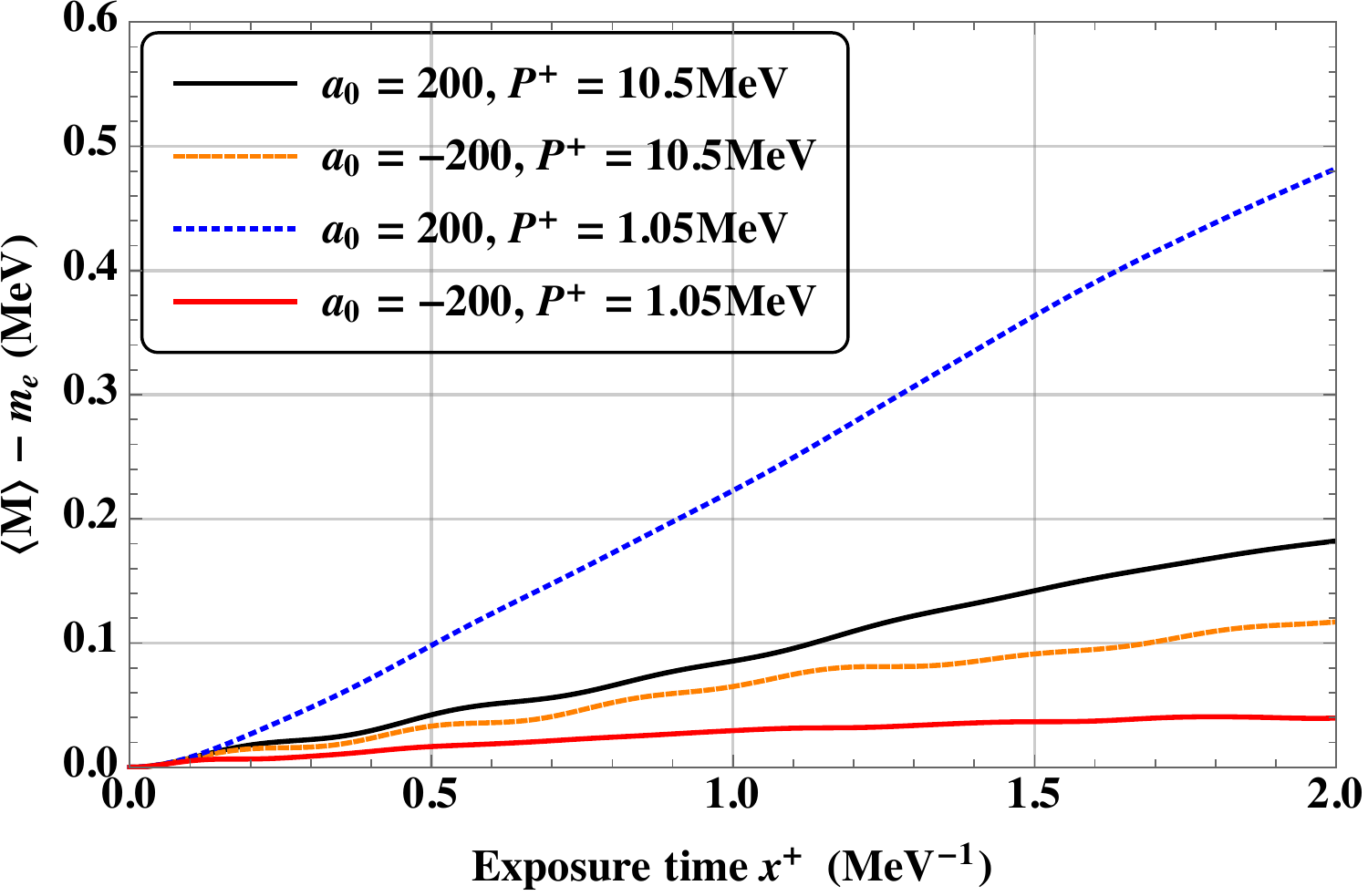} &
  \end{tabular}
  \caption{Time-evolution of the invariant mass of the physical electron in background fields (the initial invariant mass $m_{e}=0.511\rm MeV$ is subtracted). As time passes, the invariant mass is excited to higher values in all cases. Other parameters: $N_{\rm max}=100$, $K_{\rm total}=10.5$, $b_l=m_{e}$.}
   \label{fig:invmass}
   \end{center}
\end{figure*}

%
The invariant mass is defined as  $M^{2}=P^{2}=(p_{e}+p_{\gamma})^{2}$ and therefore its expectation value is 
\begin{equation}
   \langle\, M^{2}\,\rangle = \sum_{\beta}|c_{\beta}(x^{\LCp})|^2P^{\LCm}_{\beta}P^{\LCp}  \;,
   \label{invariant_mass}
\end{equation}
where $c_{\beta}(x^{\LCp})=\langle\,\beta\,|\,\Psi(x^{\LCp})\,\rangle$ is the wavefunction in the $|\,\beta\,\rangle$ basis. (Again, the transverse momentum is not included since the CM motion in the transverse directions has been eliminated.) Fig.~\ref{fig:invmass} shows the time-evolution of the expectation value of the invariant mass of the physical electron (the initial mass $m_{e}=0.511\rm MeV$ is subtracted) in the presence of background fields, with $a_{0}=\pm200$. Initially, there is only a physical electron, but as time increases a photon can be produced, which also contributes and increases the invariant mass. The increase in the invariant mass also means that energy is being injected into the system by the background.

While the invariant mass in Fig.~\ref{fig:invmass} increases for both signs of $a_{0}$, the increase is faster for $a_0>0$.  This is because, recalling that the electrons are initially centred around the origin, the repulsive background drives electrons outward toward the strong parts of the field, while in an attractive background the electrons are kept near the origin, where the field is weak. 
Again, the differences between $a_0>0$ and $a_{0}<0$ are accentuated by decreasing $P^{\LCp}$.
%
\subsubsection{Reinstating the longitudinal degrees of freedom}
%
%
As discussed in Sec.~\ref{SECT:SINGLEELECTRON}, there is no longitudinal coordinate dependence in our background field (\ref{eqn:backgroundfield}) and so the total longitudinal momentum of the system is conserved even in the presence of the background field. However, since the QED interaction couples the transverse and longitudinal degrees of freedom, the relative fraction of the total (conserved) longitudinal momentum each particle carries can change as time passes. In this section we study the longitudinal degrees of freedom.

\begin{figure*}[t]
\centering
   \begin{center}
  \begin{tabular}{@{}cccc@{}}
     \includegraphics[width=.47\textwidth]{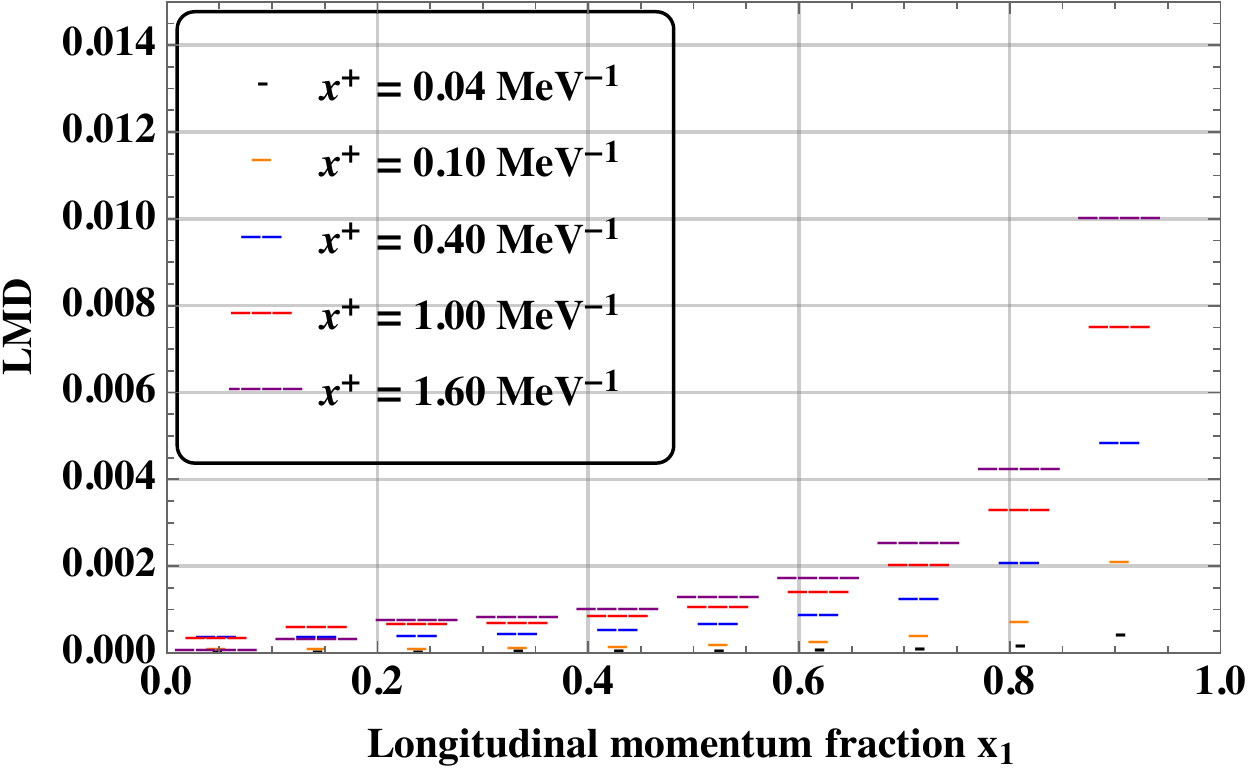} &
     \includegraphics[width=.47\textwidth]{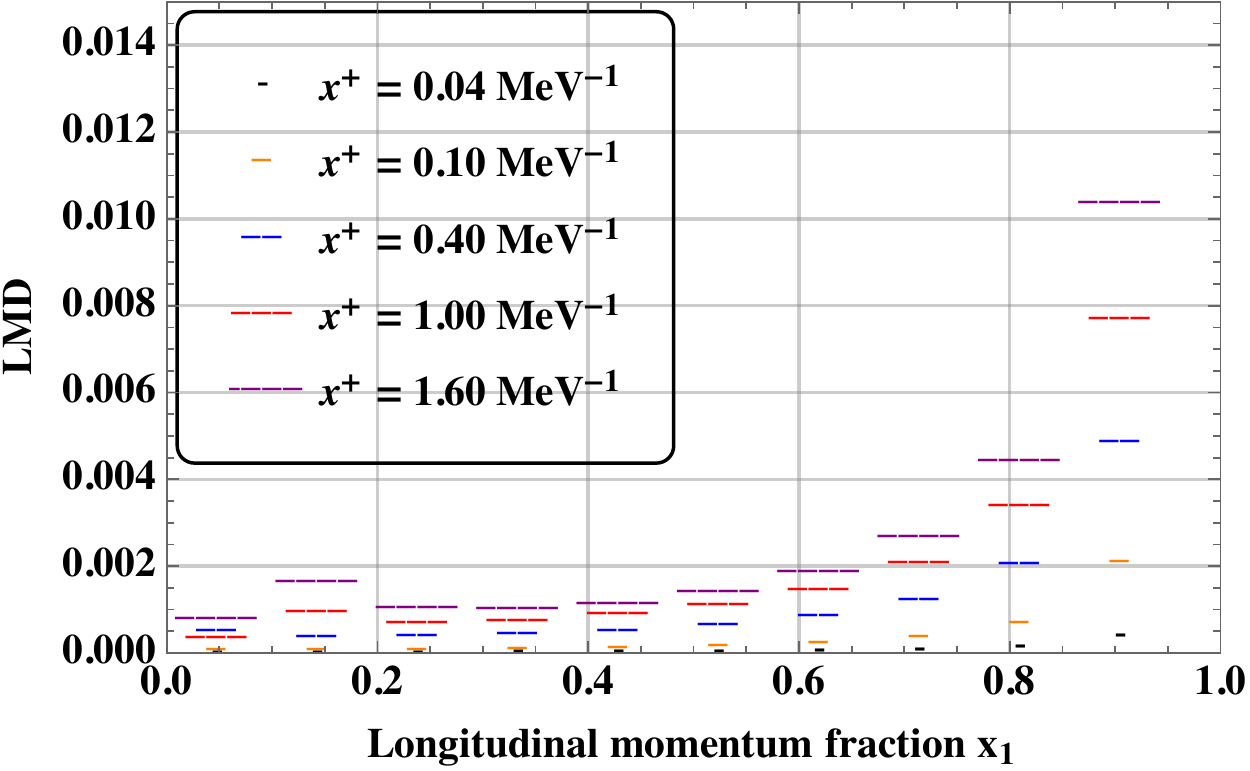} &\\
     \includegraphics[width=.47\textwidth]{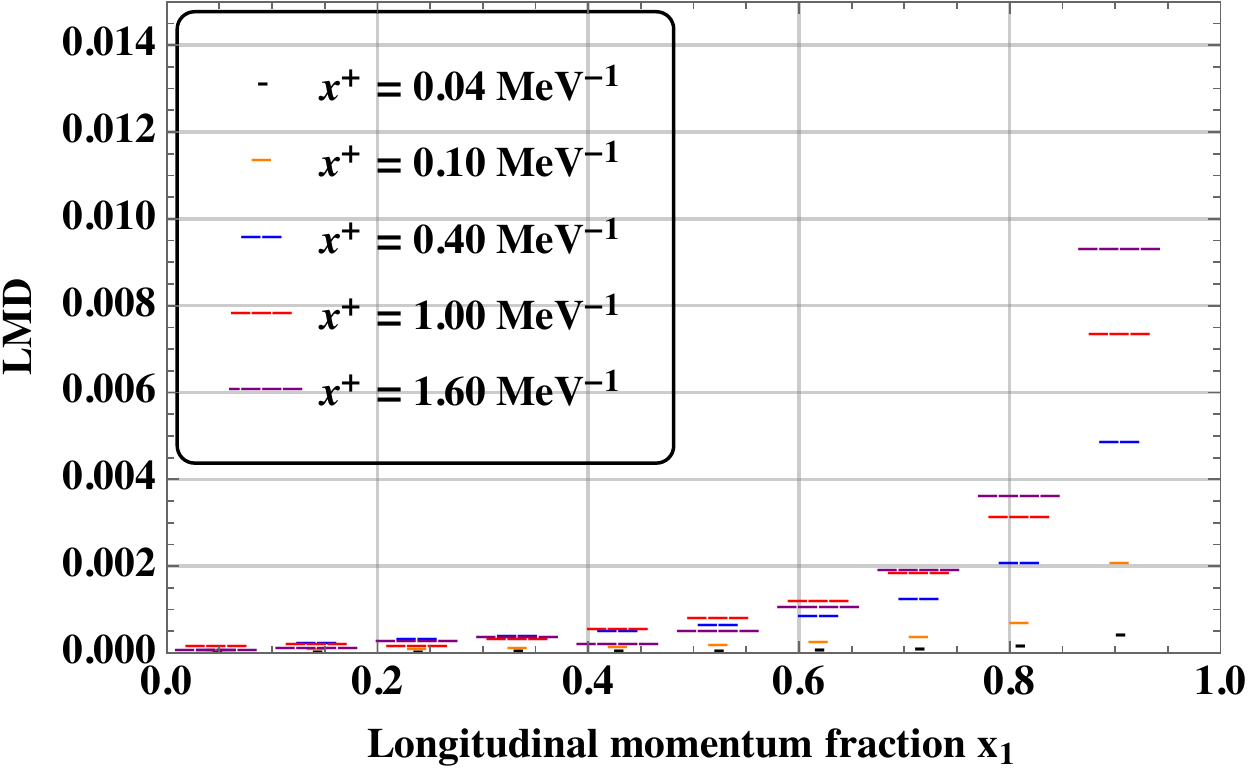} &
     \includegraphics[width=.47\textwidth]{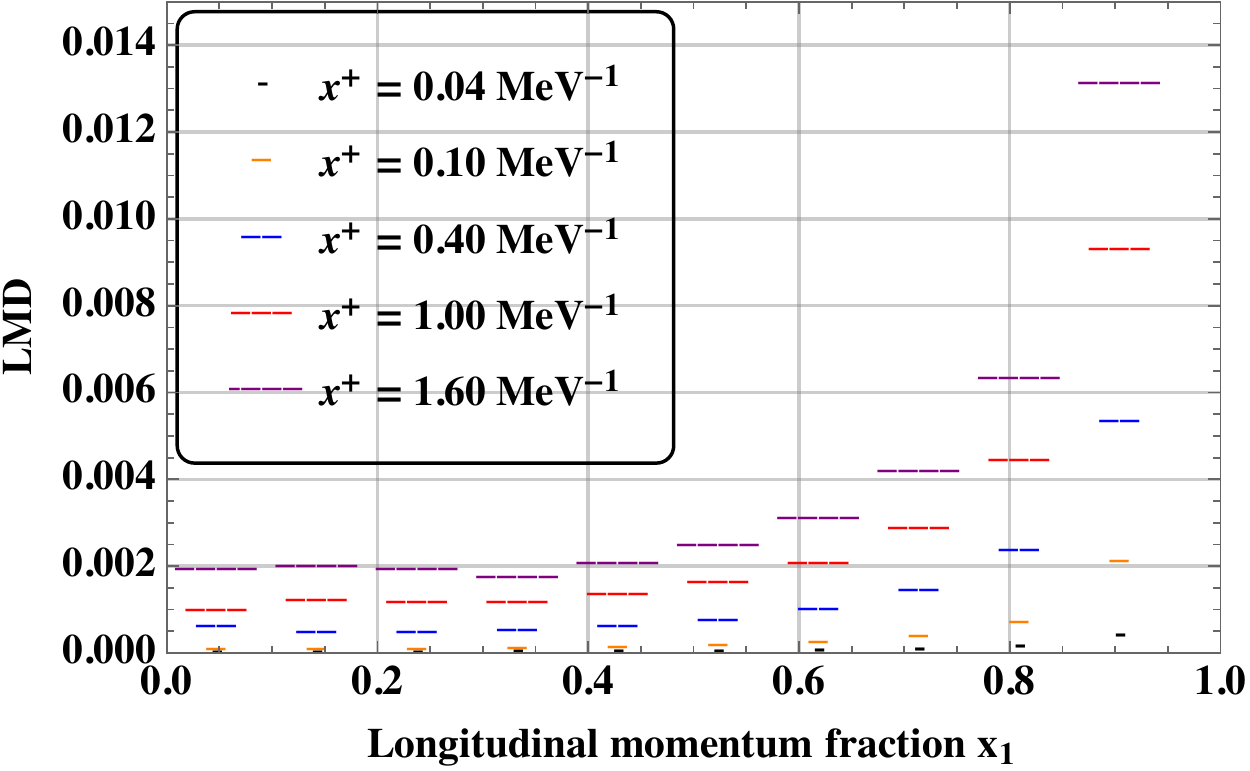} &
  \end{tabular}
  \caption{Time-evolution of the LMD. In all plots, $N_{\rm max}=100$, $K_{\rm total}=10.5$, $b_l=m_{e}$. In the left-hand column, $a_0=-200$; in the right-hand column $a_0=200$. Longitudinal momentum $P^{\LCp}$ decreases from $P^{\LCp}=10.5\rm MeV$ to $P^{\LCp}=1.05\rm MeV$ as we move from the upper to the lower row. In all the panels the distributions continue to increase with time at almost every electron longitudinal momentum fraction $\rm x_{1}$. In the left column (attractive field), as $P^{\LCp}$ decreases, the smaller $\rm x_{1}$ excitations decrease. Contrastingly, in the right column (repulsive field), the smaller $\rm x_{1}$ excitations grow as $P^{\LCp}$ decreases.}
   \label{fig:pdftimedependence}
   \end{center}
\end{figure*}

In fact, we will begin with the longitudinal degrees of freedom \textit{only}, by integrating out the transverse degrees of freedom and considering the longitudinal motion of the electron-photon state (LMD) and its dependence on the background field. Fig.~\ref{fig:pdftimedependence} shows the time-evolution of LMD in backgrounds with both $a_{0}>0$ and $a_{0}<0$ (left- and right-hand columns respectively). In the upper and lower rows we use total longitudinal momenta $P^{\LCp}=10.5\rm MeV$ and $P^{\LCp}=1.05\rm MeV$ respectively. The probability at (almost) every $\mathrm x_{1}$ increases with time in background fields with both $a_{0}>0$ and $a_{0}<0$. As shown in the left column ($a_{0}=-200$), there are fewer small $\rm x_{1}$ excitations as we reduce the total longitudinal momentum (from the upper to the lower panel). By contrast, in the right column ($a_{0}=200$), there are more low $\rm x_{1}$ excitations as we reduce the total longitudinal momentum (from the upper to the lower).  This is because the photons with higher transverse momentum (see Fig.~\ref{fig:signdependence}) tend to have higher longitudinal momenta, which results in low electron momenta, i.e.~small $\rm x_{1}$.

\begin{figure}[htp]
\centering
   \begin{center}
  \begin{tabular}{@{}cccc@{}}
     \includegraphics[width=.47\textwidth]{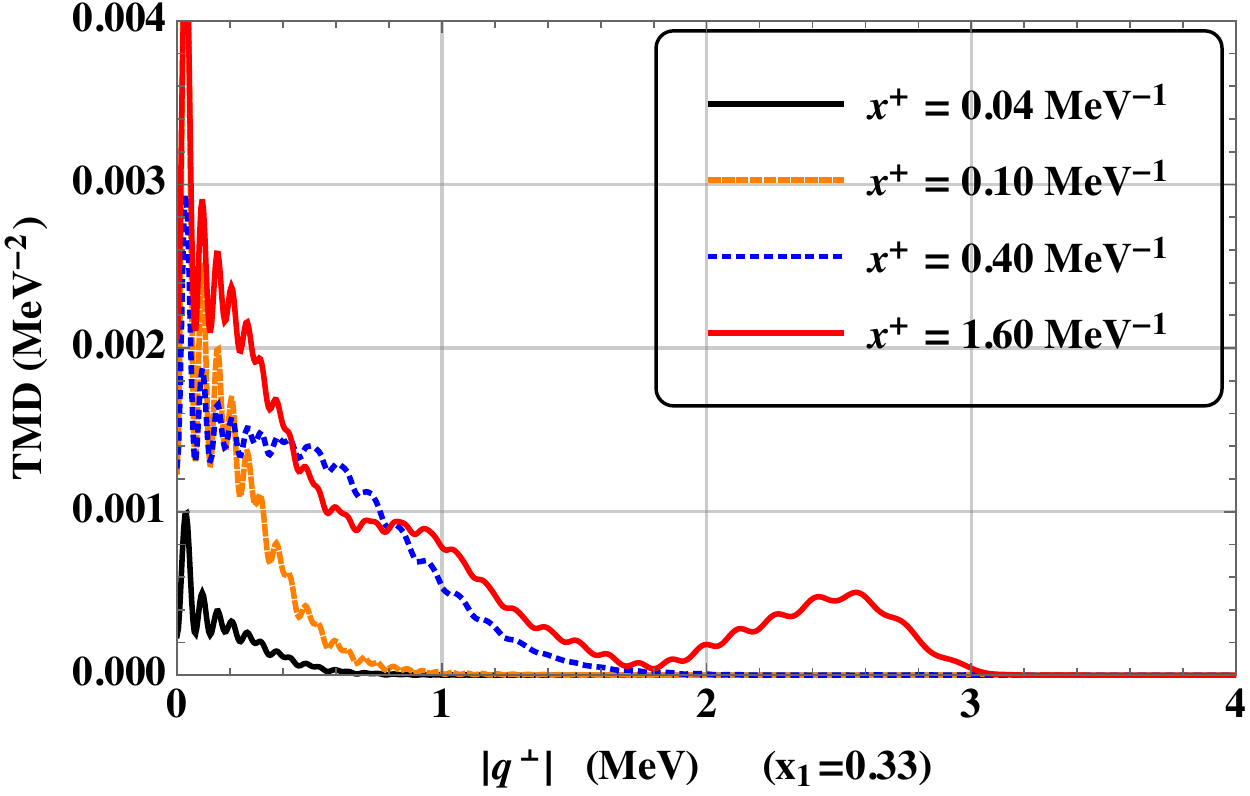} &
     \includegraphics[width=.47\textwidth]{./figures/relative/n100k10e100ks5_2.pdf} &\\
     \includegraphics[width=.47\textwidth]{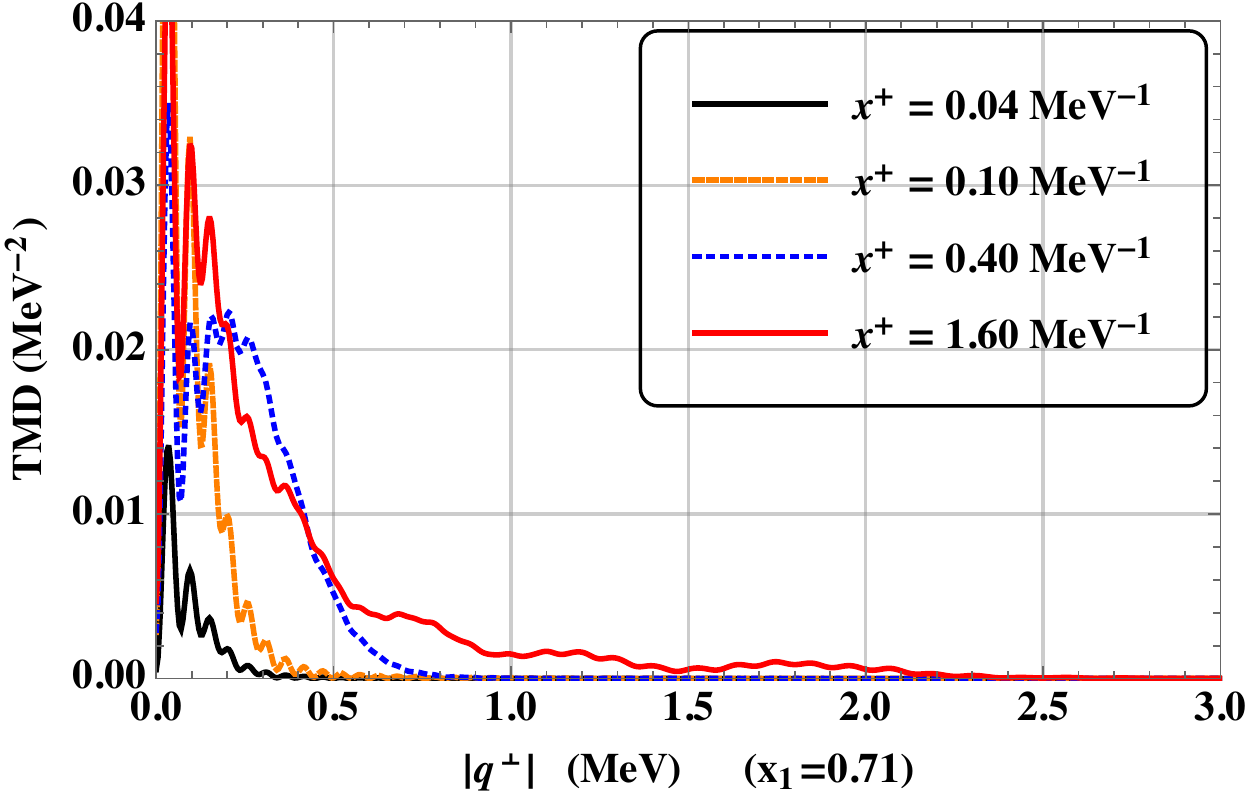} &
     \includegraphics[width=.47\textwidth]{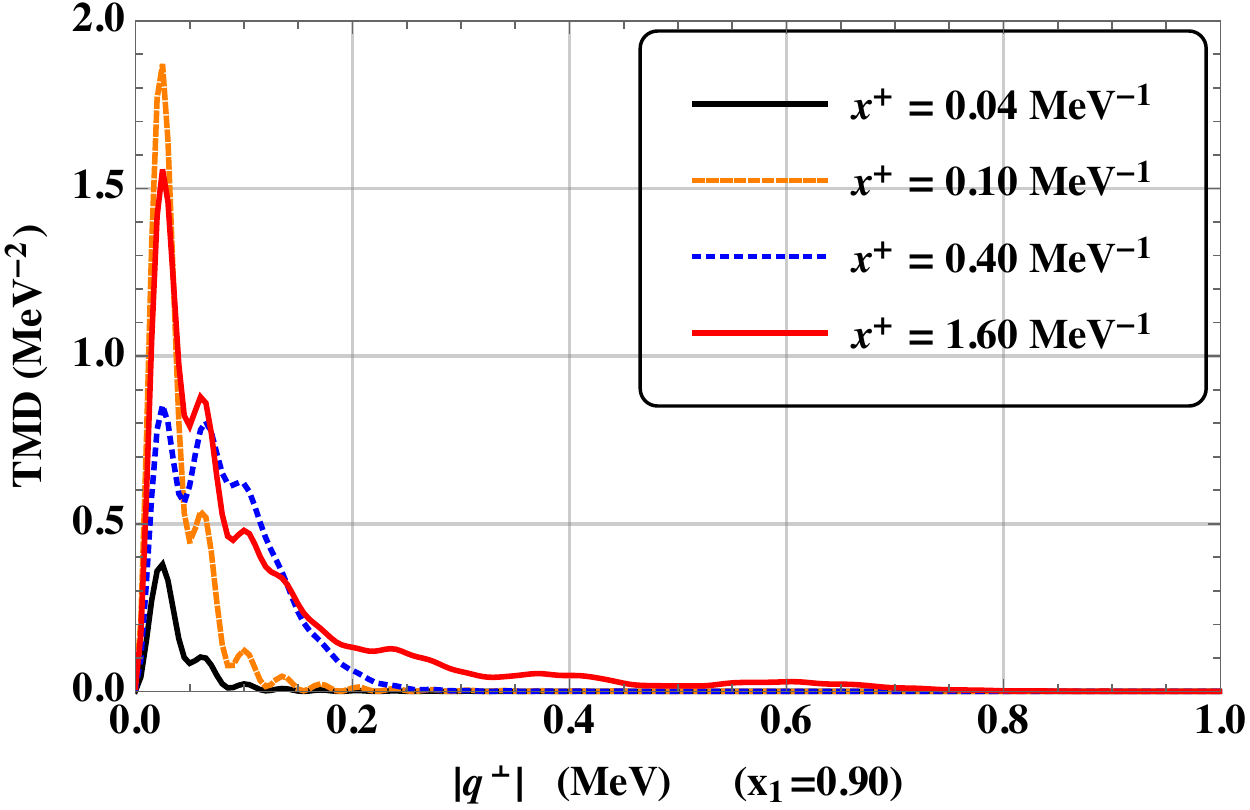} &
  \end{tabular}
  \caption{Time-evolution of the TMDs for longitudinal momentum fractions $\mathrm x_{1}=0.33, 0.52, 0.71$ and $0.90$, as indicated in each plot; the corresponding widths of the HO basis are $b_{b}=0.47m_{e}, 0.50m_{e}, 0.45m_{e}$ and $0.29m_{e}$. Other parameters: $N_{\rm max}=100$, $K_{\rm total}=10.5$, $b_l=m_{e}$, $a_0=200$, $P^{\LCp}=10.5\rm MeV$. Two obvious differences between the plots (noting the different scales on both the horizontal and vertical axes) are that the larger $\rm x_{1}$ electrons have larger probability of being excited and that the distributions are excited to higher momentum more quickly for TMDs with smaller $\mathrm x_{1}$.}
   \label{fig:x1dependece}
   \end{center}
\end{figure}

Next, we consider the time-evolution of the TMDs in four different $k_{1}=3.5, 5.5, 7.5$ and $9.5$, i.e.~four different longitudinal momentum fractions of the electron, $\rm x_{1} \approx 0.33$, $0.52$, $0.71$ and $0.90$. They are shown in Fig.~\ref{fig:x1dependece}. As time passes, basis states with higher momenta are populated. One significant difference between the four panels is that the larger $\mathrm x_{1}$ electrons have larger probability (note the scale of the y-axes) of being excited, meaning that the photons generated by the electron in the presence of the background field tend to have small longitudinal momenta. This can be explained by
\begin{equation}
   s_{2}=\frac{q^{\LCperp 2}+m_{e}^{2}}{\mathrm x_{1}}+\frac{q^{\LCperp 2}}{\mathrm x_{2}}\;,
   \label{invmass}
\end{equation}
which approximates the invariant mass of the outgoing electron-photon states. As Eq.~(\ref{invmass}) shows, because of the nonzero mass of the electron, small $\rm x_{1}$ electrons carry more energy and therefore the corresponding photons, with $\mathrm x_{2}=1-\mathrm x_{1}$, have less probability of being excited. Another main difference between the four panels in Fig.~\ref{fig:x1dependece} is that the momenta increase faster (note the scale of the x-axes) in panels with smaller longitudinal momentum fractions. This is because, as in BLFQ and perturbation theory \cite{Brodsky:1997de}, the physical electron (our initial state) has broader TMDs in smaller $\mathrm x_{1}$; in a Gaussian background field of width $b_{l}$, the width of the TMDs does not change too much and therefore the TMDs with smaller $\mathrm x_{1}$ are still wider (move faster).
%

\section{Photon Emission in Time-Dependent Background Fields} \label{SECT:TIME}
In previous sections the background field was independent of time; i.e.~we took $f\equiv 1$ in (\ref{BG-AI}). In such backgrounds, the probability that the physical electron emits a physical photon will naturally continue to rise as time evolves (as will the relative momentum between the electron and photon). In this section we add a time dependence to the background field by taking $f(x^{\LCp})=\sin(\omega x^{\LCp})$ in Eq.~(\ref{BG-AI}), where $\omega =1.92\rm{MeV}$, chosen such that the energy difference between the lowest state and the 100th state (with energies labelled $P_{1}^{\LCm}$ and $P_{100}^{\LCm}$) is $P^{\LCm}_{100}-P^{\LCm}_{1}=2\omega$. We fix $K_{\rm total}=1.5$ and $N_{\rm max}=160$, and thus the basis in the transverse plane is more complete\footnote{In this section we retain the lowest 318 states, which means that we retain the states with invariant mass less than $8.85\rm MeV$.}; as a result, the longitudinal momentum fraction can only take one value $\mathrm x_{1}=0.5/1.5\approx 0.33$. In this section, we fix $L=2 \pi \mathrm{MeV}^{-1}$ and therefore $P^{\LCp}=1.5\rm MeV$.

\begin{figure*}[t]
\centering
   \begin{center}
      \begin{tabular}{@{}cccc@{}}
     \includegraphics[width=.47\textwidth]{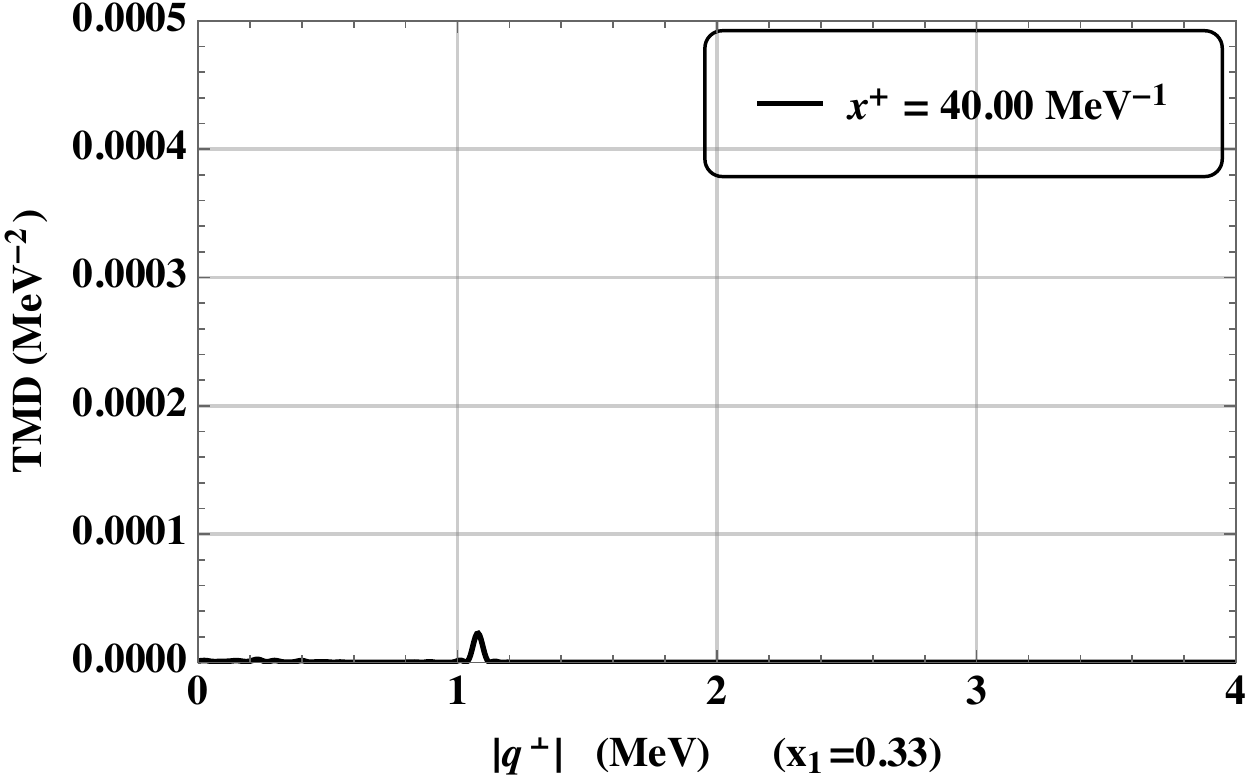} 
     \includegraphics[width=.47\textwidth]{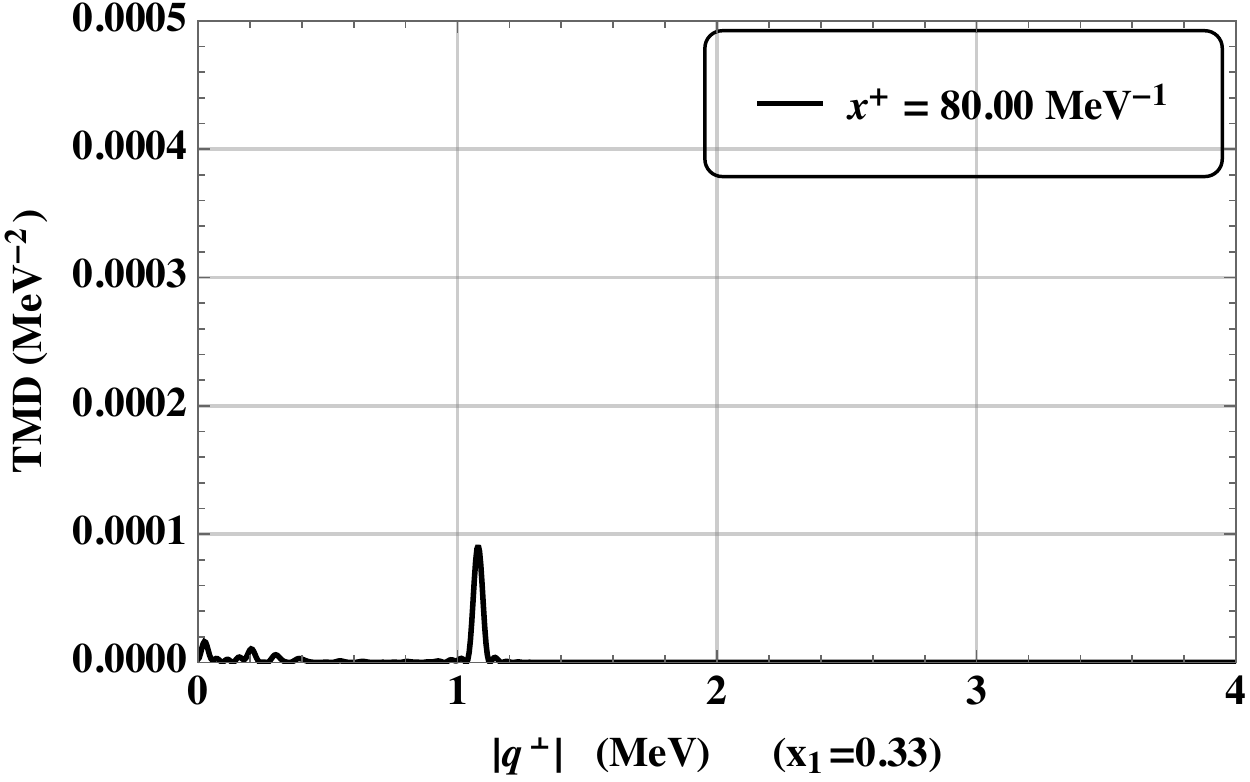} \\
     \includegraphics[width=.47\textwidth]{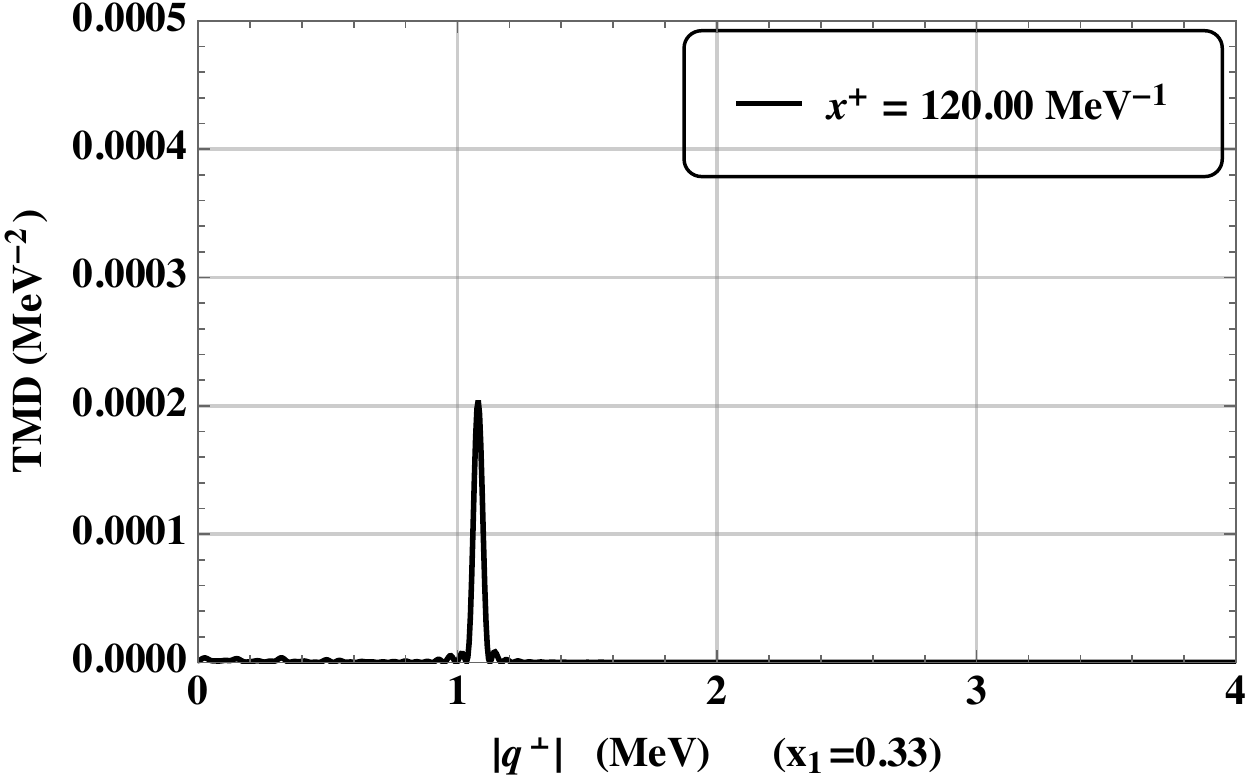} 
     \includegraphics[width=.47\textwidth]{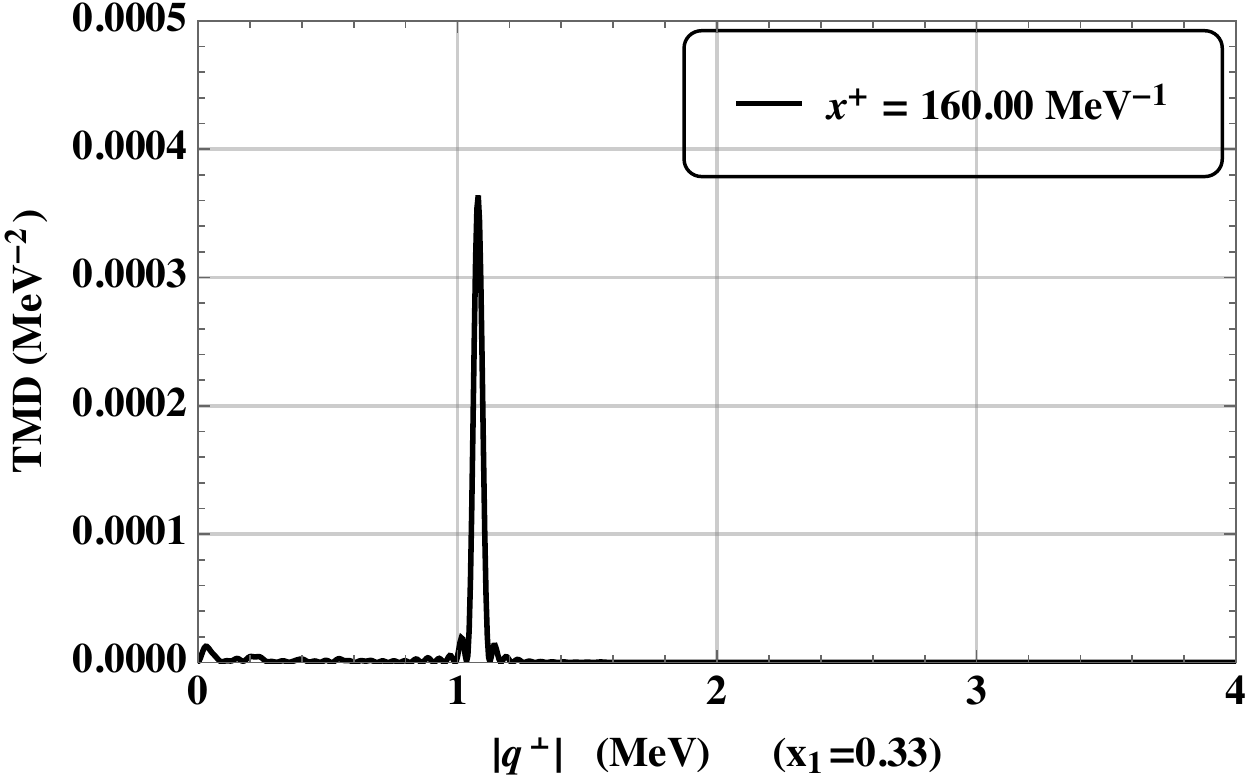} 
  \end{tabular}
  \caption{Time-evolution of TMD in a time-dependent background field $(a_{0}=1)$ with a time dependence $f(x^{\LCp})=\sin(\omega x^{\LCp})$ in which $\omega =1.92\rm{MeV}$. Four panels show TMDs in increasing light-front time $x^{\LCp}$. 
  As time passes, only the excitation from the state with $\Delta P^{\LCm}=2\omega$ increases significantly. Other parameters: $N_{\rm max}=160$, $K_{\rm total}=1.5$, $b_{b}=0.47m_{e}$, $b_l=m_{e}$, $\mathrm x_{1}=0.33$, $P^{\LCp}=1.5\rm MeV$. }
   \label{fig:timetmda1}
   \end{center}
\end{figure*}

We consider two cases, $a_{0}=1$ and $a_{0}=20$, corresponding to comparatively weak and strong backgrounds. Fig.~\ref{fig:timetmda1} shows the TMD for the case $a_0=1$. The TMD consists of, essentially, a single peak, the height of which grows as time passes. The peak is located at an energy difference, relative to the ground state, of $\Delta P^{\LCm}=P^{\LCm}_{\beta}-P^{\LCm}_{1}=2\omega$.  (Recall that the excited states in $|\,\beta\,\rangle$ are scattering states, and therefore appear as peaks in the TMD.) This agrees well with Fermi's golden rule.

\begin{figure*}[t]
\centering
   \begin{center}
      \begin{tabular}{@{}cccc@{}}
     \includegraphics[width=.47\textwidth]{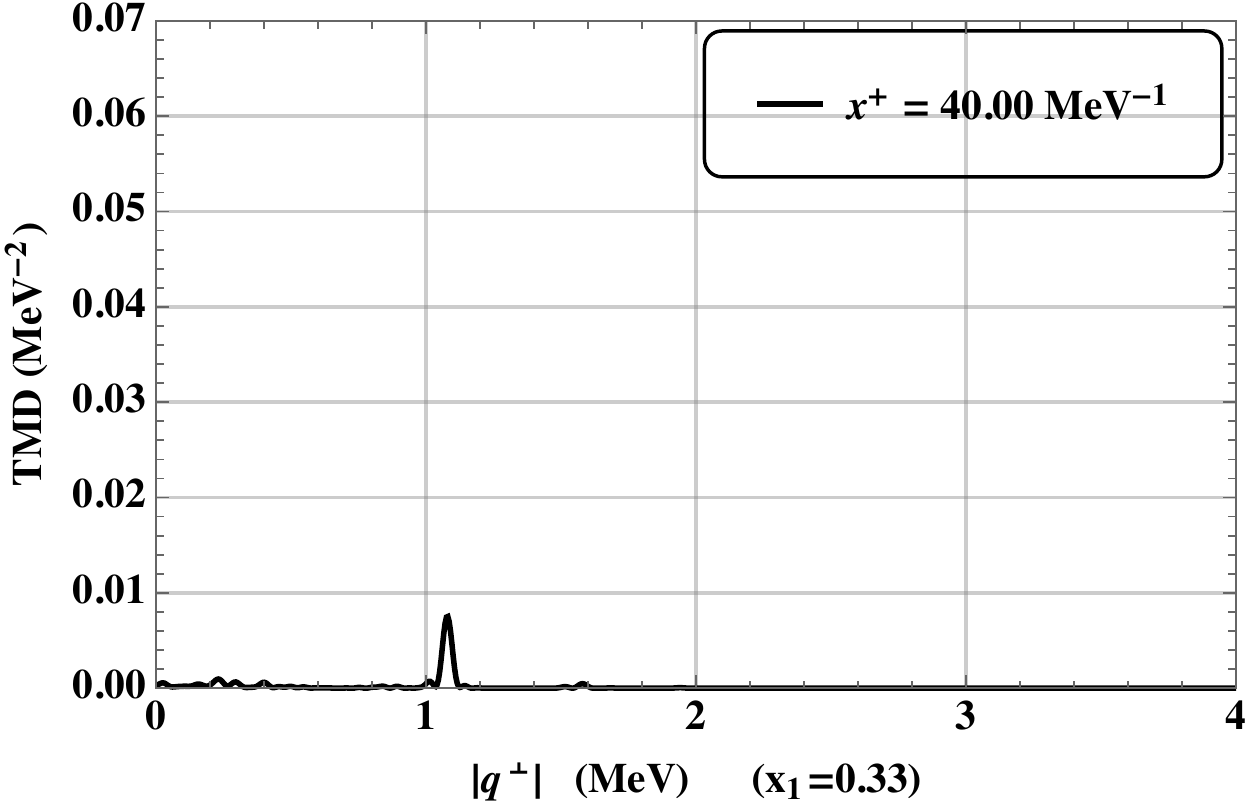} 
     \includegraphics[width=.47\textwidth]{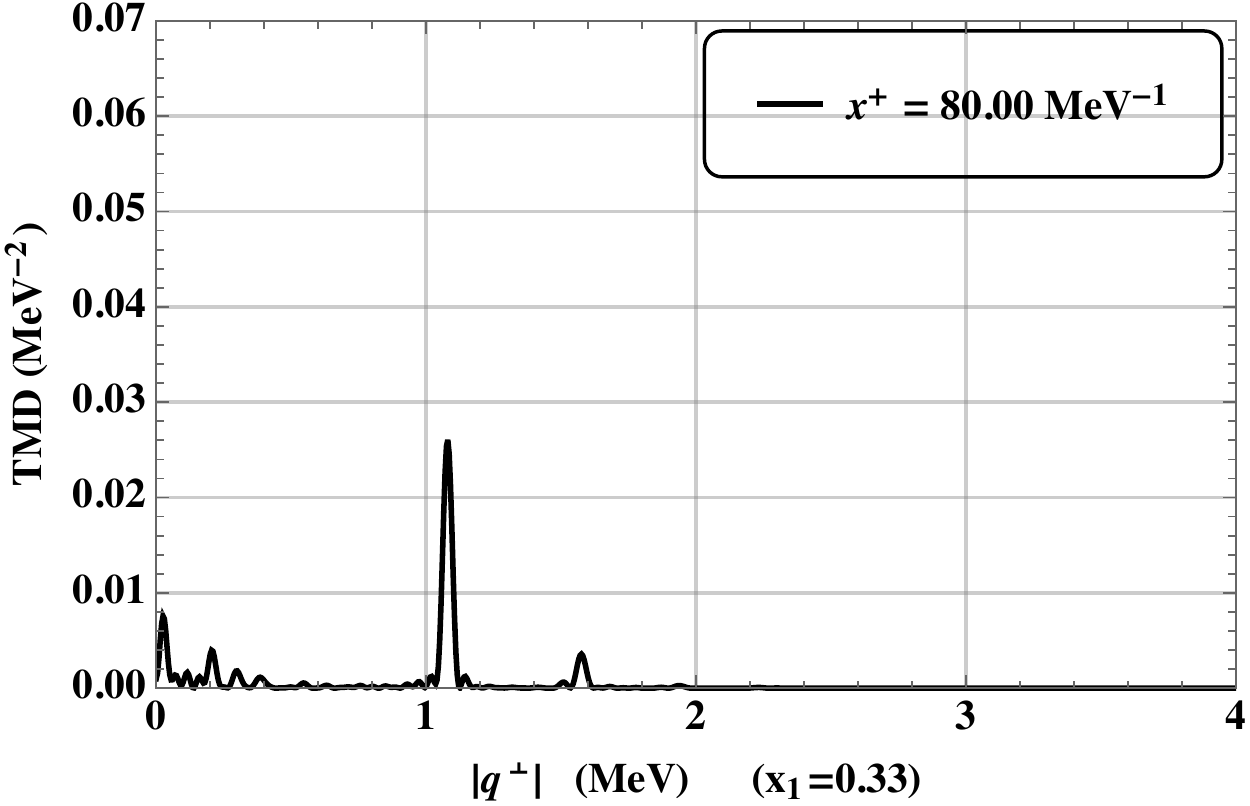} \\
     \includegraphics[width=.47\textwidth]{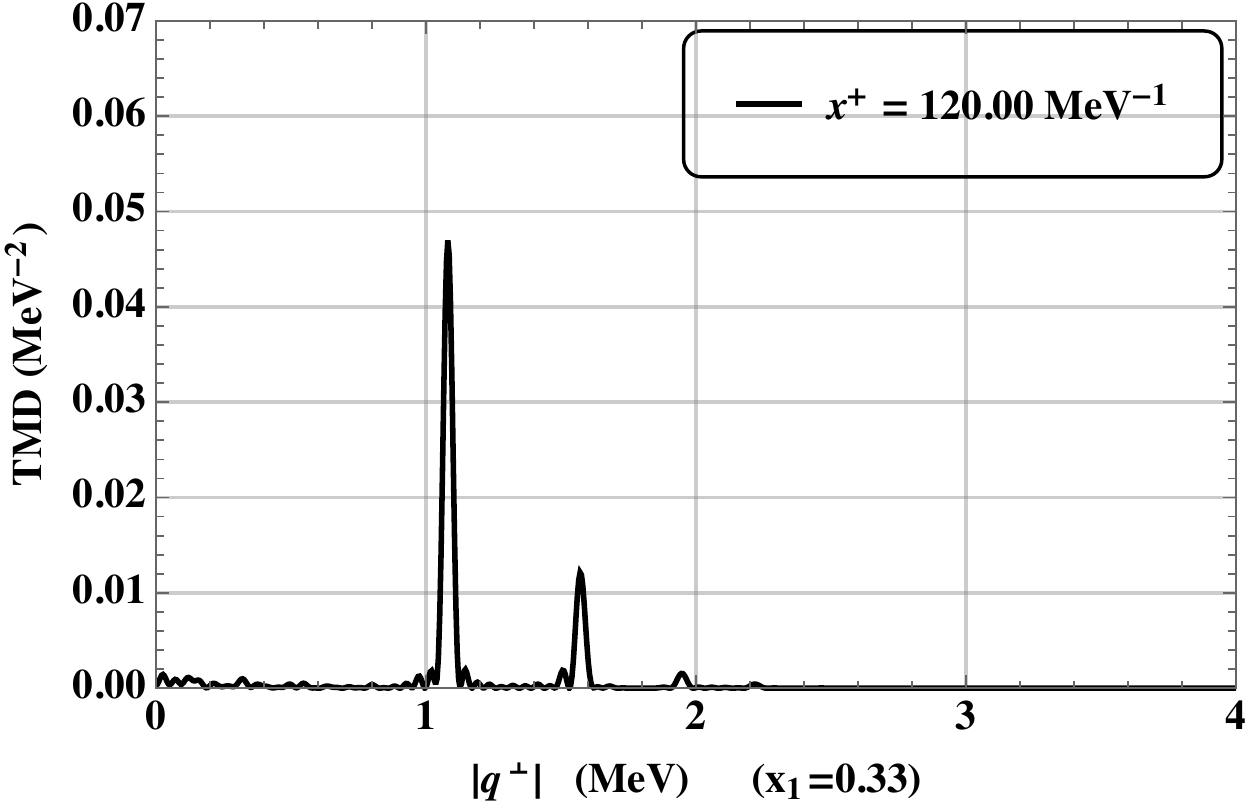} 
     \includegraphics[width=.47\textwidth]{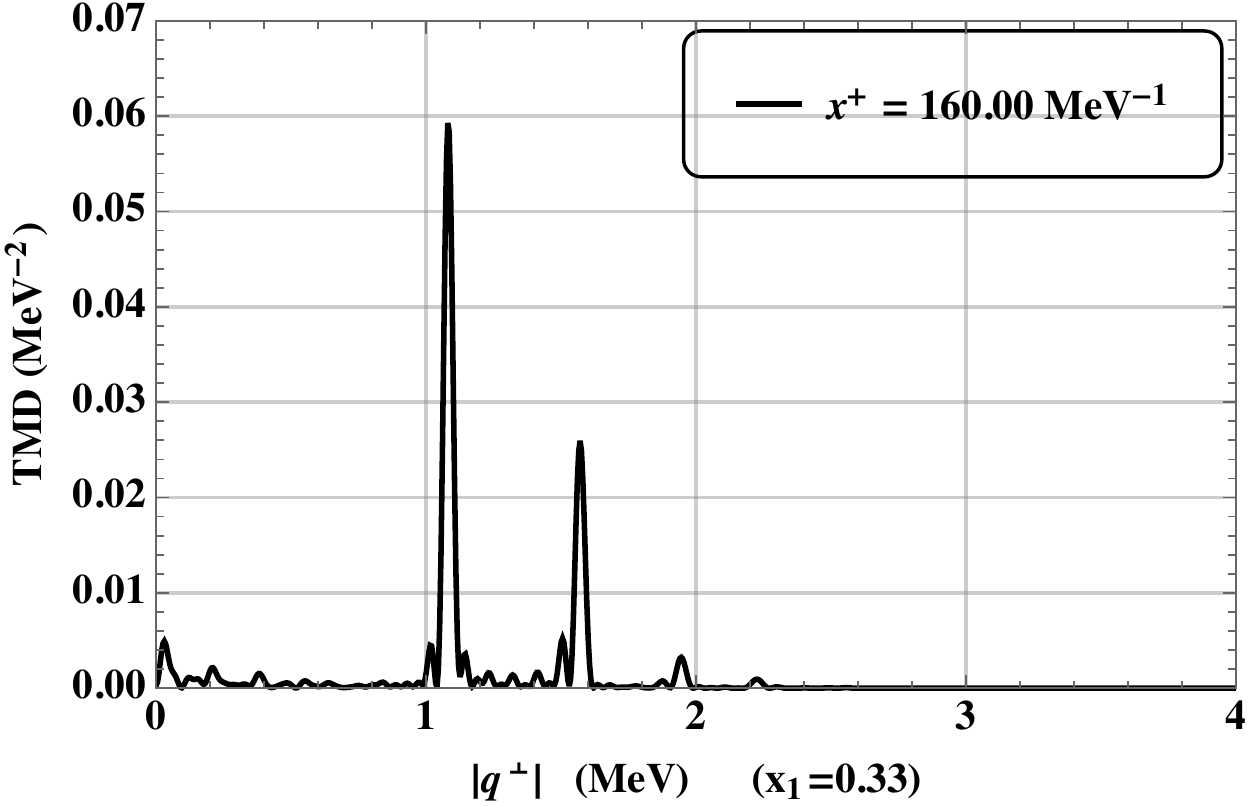} 
  \end{tabular}
  \caption{Time-evolution of TMD in a time-dependent background field $(a_{0}=20)$ with a time dependence $f(x^{\LCp})=\sin(\omega x^{\LCp})$ in which $\omega =1.92\rm{MeV}$. Four panels show TMDs in increasing light-front time $x^{\LCp}$. 
  At initial stages there is only the contribution from state with $\Delta P^{\LCm}=2\omega$, as time passes, the higher excited states increase as shown in the down two panels. Other parameters: $N_{\rm max}=160$, $K_{\rm total}=1.5$, $b_{b}=0.47m_{e}$, $b_l=m_{e}$, $\mathrm x_{1}=0.33$, $P^{\LCp}=1.5\rm MeV$. }
   \label{fig:timetmda20}
   \end{center}
\end{figure*}

We now increase the background field amplitude to $a_0=20$, and Fig.~\ref{fig:timetmda20} shows the resulting TMD.  Initially, for small elapsed time, there is only a single peak, again at an energy corresponding to $\Delta P=2\omega$, as for $a_0=1$. As time continues to pass, however, additional states are excited, and further peaks appear in the spectrum. These additional peaks are located at an energy difference of $\Delta P^{\LCm}=4\omega$, $6\omega$, $8\omega$, and so on. These can be understood as \textit{higher harmonics} of the laser field, which are excited due to the \textit{nonlinear} interaction of the background with the system.
\section{Conclusions and outlook}\label{SECT:OUTLOOK}
In this paper, we have used tBLFQ to study the dynamics of an electron interacting with an intense background field, and its consequent emission of radiation (nonlinear Compton scattering). Our chosen background was a model of an axicon laser beam, which has a nontrivial spacetime structure in the directions transverse to the beam propagation. It is by including this structure that we have been able to go beyond previous tBLFQ studies in this area. Notably, the geometrical properties of the basis used in tBLFQ seem particularly well matched with, and thus suitable for the study of, the geometry of the considered backgrounds. 

In our analysis of nonlinear Compton scattering we began with the acceleration of the electron in the background, within only the $|\,e\,\rangle$ Fock sector included in the calculation; the results agree with a sampling of the classical calculation. Then, in the study of the relative motion, we defined two useful observables. The first, the LMD, gives the probability of finding a scattered electron-photon state at some longitudinal momentum fraction. We saw that the LMD increases with time for every value of the longitudinal momentum fraction, implying that photons can be generated at all possible longitudinal momenta. The second observable, the TMD, gives the distribution of the relative momentum between the electron and the photon in the transverse plane. As time passes, the TMDs increase with the centre of the distributions moving to higher momentum; this implies that photons are being emitted from the physical electron due to the presence of the background, which naturally leads to an excitation in the relative motion between the photons and the electrons.

We have considered fields which have spacetime dependence on transverse position, and also fields which depend additionally on (light-front) time; including also a dependence on the remaining longitudinal dependence is required to make the field fully realistic, of course. This provides no real obstacle to tBLFQ, however, other than an increase in numerical complexity, and will be considered elsewhere.

Future research directions include the further development of tBLFQ itself, and the study of other processes\cite{Chen:2017uuq,Chen:2017mat}. For example, here we have included only the first two Fock sectors ($|\,e\,\rangle$ and $|\,e\gamma\,\rangle$), but in order to obtain a complete picture of e.g.~multiphoton emission in Compton scattering, more Fock sectors are needed. Other potential applications of tBLFQ include studying the time-evolution of states in the strong electromagnetic fields generated by heavy ion collisions and studying particle production (the Sauter-Schwinger effect) in strong backgrounds. Developments are also needed in the handling of the centre-of-mass (CM) motion.
In Sec.~\ref{SECT:WITHBACKGROUND}, we boosted the system to the CM frame, in which there is by definition no CM motion. Our background is invariant under this transformation, up to a displacement, which has however not been included in this paper. In future improvements, we could replace the CM motion by a classical calculation and treat the background field as a (classically) moving background field in the CM frame.

\acknowledgments
The authors thank Guangyao Chen, Meijian Li, Siqi Xu, Biaogang Wu, Kaiyu Fu and James P.~Vary for insightful discussions. The authors are supported by new faculty startup funding by the Institute of Modern Physics, Chinese Academy of Sciences and by Key Research Program of Frontier Sciences, CAS, Grant No ZDBS-LY- 7020 (X.Z.) and the EPSRC, contract EP/S010319/1 (A.I.). A portion of the computational resources was provided by the National Energy Research Scientific Computing Center (NERSC), which is supported by the US DOE Office of Science.
\appendix

\section{Light-front full Hamiltonian}
%
We start with the QED Lagrangian 
\begin{equation}
   \mathcal{L}=-\frac{1}{4} F_{\mu \nu} F^{\mu \nu}+\bar{\Psi}\left(i \gamma^{\mu} D_{\mu}-m_{e}\right) \Psi\;,
\end{equation}
in which the covariant derivative $D_{\mu} := \partial_{\mu}+i e (A_{\mu}+\mathcal{A}_{\mu})$ contains both the dynamical photon field and a background field $\mathcal{A}$. By taking the Legendre transformation, the full light-front Hamiltonian may be obtained as~\cite{Zhao:2013cma}
\begin{equation} 
   \begin{split}
      P^{\LCm}=& \int \mathrm{d}^{2} x^{\LCperp} \mathrm{d} x^{\LCm} \frac{1}{2} \bar{\Psi} \gamma^{\LCp} \frac{m_{e}^{2}+\left(i \partial^{\LCperp}\right)^{2}}{i \partial^{\LCp}} \Psi \\ &+\frac{1}{2} A^{j}\left(i \partial^{\LCperp}\right)^{2} A^{j}+e j^{\mu} A_{\mu}+\frac{e^{2}}{2} j^{\LCp} \frac{1}{\left(i \partial^{\LCp}\right)^{2}} j^{\LCp} \\ &+\frac{e^{2}}{2} \bar{\Psi} \gamma^{\mu} A_{\mu} \frac{\gamma^{\LCp}}{i \partial^{\LCp}} \gamma^{\nu} A_{\nu} \Psi \\ &+\frac{e^{2}}{2} \bar{\Psi} \gamma^{\mu} \mathcal{A}_{\mu} \frac{\gamma^{\LCp}}{i \partial^{\LCp}} \gamma^{\nu} \mathcal{A}_{\nu} \Psi+\frac{e^{2}}{2} \bar{\Psi} \gamma^{\mu} A_{\mu} \frac{\gamma^{\LCp}}{i \partial^{\LCp}} \gamma^{\nu} \mathcal{A}_{\nu} \Psi \\ &+\frac{e^{2}}{2} \bar{\Psi} \gamma^{\mu} \mathcal{A}_{\mu} \frac{\gamma^{\LCp}}{i \partial^{\LCp}} \gamma^{\nu} A_{\nu} \Psi+e j^{\mu} \mathcal{A}_{\mu}\;.
      \label{eqn:fullhamiltonian}
   \end{split}
\end{equation}
The first three lines are the QED light-front Hamiltonian, \(P_{\mathrm{QED}}^{\LCm} .\) The remaining lines contain the new terms generated by the background field are labelled as $V$.

\section{Mode expansions of the field operators}
%
We adopt the following mode expansions in the plane-wave basis for fermion and gauge boson field respectively
\begin{eqnarray}
   \Psi(\mathbf{x}) =\sum_{\lambda}\int \frac{d p^{\LCp} d^{2} p^{\LCperp}}{\sqrt{2(2 \pi)^{3}}}\left(b(\mathbf{p},\lambda) u(\mathbf{p}, \lambda) e^{-i \mathbf{p}\cdot\mathbf{x}}+d^{\dagger}(\mathbf{p},\lambda) v(\mathbf{p}, \lambda) e^{+i \mathbf{p} \cdot\mathbf{x}}\right)\;, 
   \\ {A}_{\mu}(\mathbf{x}) =\sum_{\lambda} \int \frac{d p^{\LCp} d^{2} p^{\LCperp}}{\sqrt{2p^{\LCp}(2 \pi)^{3}}}\left(a(\mathbf{p},\lambda) \epsilon_{\mu}(\mathbf{p}, \lambda) e^{-i \mathbf{p} \cdot\mathbf{x}}+a^{\dagger}(\mathbf{p},\lambda) \epsilon_{\mu}^{*}(\mathbf{p}, \lambda) e^{+i \mathbf{p}\cdot\mathbf{x}}\right)\;, 
\end{eqnarray}
where \(\mathbf{p} \cdot \mathbf{x}:=\frac{1}{2} p^{\LCp} x^{\LCm}-p^{\LCperp} \cdot x^{\LCperp}\) is the $3$-product for the spatial components of \(p^{\mu}\) and \(x^{\mu}\). The creation and annihilation operators obey the (anti)commutation relations
\begin{equation}
   \left[{a}(\mathbf{p},\lambda), {a}^{\dagger}\left(\mathbf{p'},\lambda'\right)\right]=\left\{{b}(\mathbf{p},\lambda), {b}^{\dagger}\left(\mathbf{p'},\lambda'\right)\right\}=\left\{{d}(\mathbf{p},\lambda), d^{\dagger}\left(\mathbf{p'},\lambda'\right)\right\}=\delta\left(p^{\LCp}-p'^{\LCp}\right) \delta^{(2)}\left({p}^{\LCperp}-{p'}^{\LCperp}\right) \delta_{\lambda}^{\lambda'}\;.
   \label{eqn:commutation}
\end{equation}
The single-particle states are
   \begin{equation}
      |\,\mathbf{p},\lambda\,\rangle_{e} \equiv b^{\dagger}(\mathbf{p},\lambda)|\,0\,\rangle \text{~and~} |\,\mathbf{p},\lambda\,\rangle_{\gamma}\equiv a^{\dagger}(\mathbf{p},\lambda)|\,0\,\rangle\;.
      \label{}
   \end{equation}
The Dirac spinors are
\begin{equation}
   \begin{split}
      u(\mathbf{p}, \uparrow)=\left(\begin{array}{c}{1} \\ {0} \\ {\frac{i m_{e}}{p^{\LCp}}} \\ {\frac{\left(i p^{1}-p^{2}\right)}{p^{\LCp}}}\end{array}\right)\;, \quad u(\mathbf{p}, \downarrow)=\left(\begin{array}{c}{0} \\ {1} \\ {\frac{\left(-i p^{1}-p^{2}\right)}{p^{\LCp}}} \\ {\frac{i m_{e}}{p^{\LCp}}}\end{array}\right)\;,\\
      v(\mathbf{p}, \uparrow)=\left(\begin{array}{c}{0} \\ {1} \\ {\frac{\left(-i p^{1}-p^{2}\right)}{p^{\LCp}}} \\ {\frac{-i m_{e}}{p^{\LCp}}}\end{array}\right)\;, \quad v(\mathbf{p}, \downarrow)=\left(\begin{array}{c}{1} \\ {0} \\ {\frac{-i m_{e}}{p^{\LCp}}} \\ {\frac{\left(i p^{1}-p^{2}\right)}{p^{\LCp}}}\end{array}\right)\;.
   \end{split}
\end{equation}
The photon polarisation vectors are 
\begin{equation}
   \epsilon^{\mu}(\mathbf{k}, \lambda)=\left(0, \epsilon^{\LCperp}(\lambda), \frac{2 \epsilon^{\LCperp}(\lambda) \cdot k^{\LCperp}}{k^{\LCp}}\right)\;,
\end{equation}
where $\epsilon^{\LCperp}(\uparrow)=\frac{1}{\sqrt{2}}(1,i)$ and $\epsilon^{\LCperp}(\downarrow)=\frac{1}{\sqrt{2}}(1,-i)$. 

In a space with a compactified longitudinal direction $x^{\LCm}$ of length $2L$, as in BLFQ, the field operators become 
\begin{equation}
   \Psi(\mathbf{x}) =\sum_{k,\lambda} \int \frac{ d^{2} p^{\LCperp}}{\sqrt{2L(2 \pi)^{2}}}\left(b(\mathbf{p},\lambda) u(\mathbf{p}, \lambda) e^{-i \mathbf{p}\cdot\mathbf{x}}+d^{\dagger}(\mathbf{p},\lambda) v(\mathbf{p}, \lambda) e^{+i \mathbf{p} \cdot\mathbf{x}}\right)\;, 
\end{equation}
and 
\begin{equation}
   \\ {A}_{\mu}(\mathbf{x}) =\sum_{k,\lambda} \int \frac{d^{2} p^{\LCperp}}{\sqrt{2Lp^{\LCp}(2 \pi)^{2}}}\left(a(\mathbf{p},\lambda) \epsilon_{\mu}(\mathbf{p}, \lambda) e^{-i \mathbf{p} \cdot\mathbf{x}}+a^{\dagger}(\mathbf{p},\lambda) \epsilon_{\mu}^{*}(\mathbf{p}, \lambda) e^{+i \mathbf{p}\cdot\mathbf{x}}\right)\;, 
\end{equation}
where $p^{\LCp}=\frac{2\pi}{L}k$ with the dimensionless quantity \(k=1,2,3, \ldots\) for bosons and \(k=\frac{1}{2}, \frac{3}{2}, \frac{5}{2}, \ldots\) for fermions. Meanwhile, the Dirac delta functions of longitudinal momentum in (\ref{eqn:commutation}) should be replaced by the Kronecker delta functions
\begin{equation}
   \left[{a}(\mathbf{p},\lambda), {a}^{\dagger}\left(\mathbf{p'},\lambda'\right)\right]=\left\{{b}(\mathbf{p},\lambda), {b}^{\dagger}\left(\mathbf{p'},\lambda'\right)\right\}=\left\{{d}(\mathbf{p},\lambda), d^{\dagger}\left(\mathbf{p'},\lambda'\right)\right\}=\delta^{p'^{\LCp}}_{p^{\LCp}} \delta^{(2)}\left({p}^{\LCperp}-{p'}^{\LCperp}\right) \delta_{\lambda}^{\lambda'}\;.
   \label{eqn:dis_commutation}
\end{equation}
\section{BLFQ harmonic oscillator basis}\label{SECT:HOBASIS}
In the transverse plane, the BLFQ basis adopts the 2D-HO states, which are the eigenstates of the 2D-HO Hamiltonian
\begin{align}
    \label{HO_hami}
    H_{\rm HO}=\frac{(p^\LCperp)^{2}}{2M}+\frac{1}{2}M\Omega^2 (x^\LCperp)^{2}\;,
\end{align}
where $M$ and $\Omega$ are the mass and the frequency of the oscillator respectively. In coordinate space, the eigenfunction can be factorised into a radial and an angular function
\begin{align}
    \label{transverse_HO_wavefunction}
    \Phi^b_{nm}(\rho,\phi)=(-1)^{n}i^{|\,m|\,}f^{b}_{nm}(\rho)\chi_m(\phi) \;.
\end{align}
The corresponding eigenvalue is $E_{n,m}=(2n+|\,m|\,+1)\Omega$, in which $n$ and $m$ are integers; $n\geq0$ characterises the radial excitation and $m$ describes the angular momentum of the oscillator; \((\rho, \phi)\) are polar coordinates in the transverse plane with \(x^{1}=\rho \cos \phi\) and \(x^{2}=\rho \sin \phi .\) The explicit form of the radial function is 
\begin{align}
    f^b_{nm}(\rho)=& b\sqrt{2}\sqrt{\frac{n!}{(n+|\,m|\,)!}}\ e^{-b^2\rho^2/2} (b\rho)^{|\,m|\,}L^{|\,m|\,}_n(b^2\rho^2)\; ,
    \label{radial_HO_wavefunction}
\end{align}
where \(L_{n}^{|\,m|\,}\left(b^{2} \rho^{2}\right)\) is the generalised Laguerre polynomial and $b:=\sqrt{M\Omega}$ is the 2D-HO scale parameter. 
The angular function is
\begin{align}
   \chi_m(\phi)=\frac{1}{\sqrt{2\pi}}e^{im\phi}\;.
\end{align}
The 2D-HO eigenfunction (\ref{transverse_HO_wavefunction}) is orthonormalised such that
\begin{align}
    \int^\infty_0\int^{2\pi}_0\! \ud\rho\rho\, \ud \phi\ \Phi^{b*}_{nm}(\rho,\phi)\Phi^b_{n'm'}(\rho,\phi)\nonumber=\delta^{n'}_{n}\delta^{m'}_{m} \;.
\end{align}
The momentum-space eigenfunction can be obtained by a Fourier transform
\begin{align}
    \label{HO_wavefunction_fourier_transform}
    \nonumber \tilde\Phi^b_{nm}(p^\LCperp) &= \int\! \ud^2 x^\LCperp\ e^{-i\vec{x}^\LCperp\cdot\vec{p}^\LCperp}\Phi^{b}_{nm}(x^\LCperp) \\
    &=(2\pi)\tilde f^b_{nm}(p)\tilde\chi_m(\phi)\; ,
\end{align}
in which
\begin{align}
    \tilde f^b_{nm}(p)=&\frac{\sqrt{2}}{b}\sqrt{\frac{n!}{(n+|\,m|\,)!}}e^{-p^2/(2b^2)} \left(\frac{p}{b}\right)^{|\,m|\,}L^{|\,m|\,}_n\left(\frac{p^2}{b^2}\right)\;,
\end{align}
and
\begin{align}
    \tilde\chi_m(\phi)=\frac{1}{\sqrt{2\pi}}e^{im\phi}\;.
\end{align}

\section{Matrix elements of the interaction Hamiltonian of the background field}\label{SECT:INTMATRIX}
As in Eq.~(\ref{eqn:backgroundfield}), only the longitudinal component of the background field $\mathcal A$ is nonzero. From Eq.~(\ref{eqn:fullhamiltonian}), the only surviving term in the Hamiltonian involving the background is the last term
\begin{equation}
   V=\int_{}^{}dx^{\LCm}d^{2}x^{\LCperp}ej_{\LCm}\mathcal{A}^{\LCm}\;,
   \label{background_interaction}
\end{equation}
where $j_{\LCm}=\bar\Psi\gamma_{\LCm}\Psi=\frac{1}{2}\bar\Psi\gamma^{\LCp}\Psi$ is the fermion current. Note that the background field is treated as a classical field; therefore the photons are spectators and the nontrivial part of the interaction comes from the fermion fields. For now, we neglect the photon part and keep in mind that the matrix elements of the photon part contribute Kronecker delta functions conserving all the 4 quantum numbers for the spectator photons. In the momentum representation, the matrix elements of the interaction Hamiltonian can be straightforwardly obtained by using the commutation relations (\ref{eqn:dis_commutation})
\begin{equation}
   \begin{split}
      \langle\, p'^{\LCperp},p'^{\LCp},\lambda'\,|\,V|\,p^{\LCperp},p^{\LCp},\lambda\,\rangle
      &=\int dx^{\LCm}d^2x^{\LCperp}\frac{1}{2}e\mathcal{A^{\LCm}}(x)\langle\, p'^{\LCperp},p'^{\LCp},\lambda'\,|\,\bar\Psi(\mathbf{x})\gamma^{\LCp}\Psi(\mathbf{x})\,|\,p^{\LCperp},p^{\LCp},\lambda\,\rangle\\
     &=\int dx^{\LCm}d^2x^{\LCperp}\frac{m_ea_0}{2b_l}\sum_{ss'll'}\int \frac{d^2q'^{\LCperp} d^2q^{\LCperp}}{2L(2\pi)^{4}}\Phi^{{b_{l}}}_{00}(x^{\LCperp})
    e^{i\mathbf{(q'-q)\cdot x}}\\
    &\times\bar{u}(\mathbf{q'},s'){\gamma ^{\LCp}}u(\mathbf{q},s)\langle\, p'^{\LCperp},p'^{\LCp},\lambda'\,|\,b^{\dagger}(\mathbf{q'},s')b(\mathbf{q},s)\,|\,p^{\LCperp},p^{\LCp},\lambda\,\rangle\\
    &=\frac{m_ea_0}{b_l}\delta^{k'}_{k}\delta_{\lambda}^{\lambda'}\tilde\Phi_{00}^{b_{l}}(p'^{\LCperp}-p^{\LCperp})\;.
      \label{eqn:mome_intmat}
   \end{split}
\end{equation}
In order to transform the interaction Hamiltonian to the BLFQ basis which adopts HO eigenstates in the transverse plane, we convolute (\ref{eqn:mome_intmat}) with the HO wavefunctions
\begin{equation}
\begin{split}
   \langle\, \bar\alpha_{e}'\,|\,V\,|\,\bar\alpha_{e}\,\rangle=&
   \int \frac{d^{2}p'^{\LCperp}d^{2}p^{\LCperp}}{(2\pi)^4} \left(\tilde\Phi^{\sqrt{\mathrm{x}'}b}_{n'm'}(p'^{\LCperp})\right)^{\ast}\langle\, p'^{\LCperp},p'^{\LCp},\lambda'\,|\,V\,|\,p^{\LCperp},p^{\LCp},\lambda\,\rangle \tilde \Phi^{\sqrt{\mathrm{x}}b}_{nm}(p^{\LCperp})\\
=&\frac{m_ea_0}{b_l}\delta^{k'}_{k}\delta_{\lambda}^{\lambda'}\int\frac{d^2p'^{\LCperp} d^2p^{\LCperp}}{(2\pi)^{4}}\tilde \Phi^{{b_{l}}}_{00}(p'^{\LCperp}-p^{\LCperp})\left(\tilde \Phi^{\sqrt{\mathrm{x}'}b}_{n'm'}(p'^{\LCperp})\right)^{\ast}\tilde \Phi^{\sqrt{\mathrm{x}}b}_{nm}(p^{\LCperp})\;,
      \label{eqn:ho_intmat}
\end{split}
\end{equation}
in which the $\rm x$ (longitudinal momentum fraction) dependent HO basis is adopted, for exact factorisation in a finite truncated BLFQ basis.

We now recover the photon part: for the matrix elements in the $|\,e\gamma\,\rangle$ sector, we augment (\ref{eqn:ho_intmat}) with the Kronecker delta functions conserving photon quantum numbers and the matrix elements become
\begin{equation}
   \langle\, \bar\alpha_{e\gamma}'\,|\,V\,|\,\bar\alpha_{e\gamma}\,\rangle=\langle\, \bar\alpha_{e}'\,|\,V\,|\,\bar\alpha_{e}\,\rangle\delta_{\bar\alpha_{\gamma}}^{\bar\alpha'_{\gamma}}\;.
   \label{}
\end{equation}
Because there is no cross term between the $|\,e\,\rangle$ and $|\,e\gamma\,\rangle$ sectors, the full interaction Hamiltonian matrix is the direct sum of the interaction Hamiltonian matrix in these two sectors.
Finally, the interaction Hamiltonian matrix in BLFQ basis $|\,\alpha\,\rangle$ can be transformed to the tBLFQ basis $|\,\beta\,\rangle$, using the wavefunction $\langle\, \alpha\,|\,\beta\,\rangle$ obtained by diagonalising $P^{-}_{\rm QED}$~\cite{Zhao:2014xaa}
\begin{equation}
   \langle\, \beta'\,|\,V\,|\,\beta\,\rangle=\sum_{\alpha\alpha'}\langle\, \beta'\,|\,\alpha'\,\rangle\langle\, \alpha'\,|\,V\,|\,\alpha\,\rangle\langle\, \alpha\,|\,\beta\,\rangle \;.
   \label{}
\end{equation}



\bibliography{article} 



  



\end{document}